\documentclass{emulateapj}

\slugcomment{Draft of \today}

\shorttitle{SN~2010da}
\shortauthors{Lau et al.}

\usepackage{amsmath}
\usepackage{graphicx}
\usepackage[section] {placeins}
\usepackage{hyperref}
\usepackage{ulem}

\newcommand{\beq}{\begin{equation}}
\newcommand{\eeq}{\end{equation}}

\begin{document}

\title{Rising from the Ashes: Mid-Infrared Re-Brightening of the Impostor SN~2010da in NGC~300\footnote{
Based in part on observations with the 1.3-m telescope operated by the SMARTS
Consortium at Cerro Tololo Inter-American Observatory.}}

\author{Ryan M. Lau\altaffilmark{1,2},
Mansi M. Kasliwal\altaffilmark{2},
Howard E. Bond\altaffilmark{3,4},
Nathan Smith\altaffilmark{5},
Ori D. Fox\altaffilmark{4},
Robert Carlon\altaffilmark{6},
Ann Marie Cody\altaffilmark{7},
Carlos Contreras\altaffilmark{8},
Devin Dykhoff\altaffilmark{6},
Robert Gehrz\altaffilmark{6},
Eric Hsiao\altaffilmark{9},
Jacob Jencson\altaffilmark{2},
Rubab Khan\altaffilmark{10},
Frank Masci\altaffilmark{11},
L. A. G. Monard\altaffilmark{12},
Andrew J. Monson\altaffilmark{3},
Nidia Morrell\altaffilmark{13},
Mark Phillips \altaffilmark{13},
Michael E. Ressler\altaffilmark{1}
}

\altaffiltext{1}{Jet Propulsion Laboratory, California Institute of Technology, 4800 Oak Grove Drive, Pasadena, CA 91109, USA}
\altaffiltext{2}{California Institute of Technology, Pasadena, CA 91125, USA}
\altaffiltext{3}{Dept. of Astronomy \& Astrophysics, Pennsylvania State University, University
Park, PA 16802 USA}
\altaffiltext{4}{Space Telescope Science Institute, 3700 San Martin Dr., Baltimore, MD 21218 USA}
\altaffiltext{5}{Steward Observatory, University of Arizona, Tuscon, AZ 85721, USA}
\altaffiltext{6}{Minnesota Institute for Astrophysics, School of Physics and Astronomy, 116 Church Street, S. E., University of Minnesota, Minneapolis, MN 55455, USA}
\altaffiltext{7}{NASA Ames Research Center, Moffett Field, CA 94035, USA}
\altaffiltext{8}{Las Campanas Observatory, Carnegie Observatories, Casilla
601, La Serena, Chile}
\altaffiltext{9}{Department of Physics, Florida State University, 77 Chieftain Way, Tallahassee, FL, 32306, USA}
\altaffiltext{10}{NASA Goddard Space Flight Center, MC 665, 8800 Greenbelt Road, Greenbelt, MD 20771}
\altaffiltext{11}{Infrared Processing and Analysis Center, California Institute of Technology, M/S 100-22, Pasadena, CA 91125, USA}
\altaffiltext{12}{Bronberg and Kleinkaroo Observatories, PO Box 281, Calitzdorp 6660 Western Cape, South Africa}\
\altaffiltext{13}{Carnegie Institution of Washington, Las Campanas Observatory, Colina el Pino, Casilla 601, La Serena, Chile}

\begin{abstract}

We present multi-epoch mid-infrared (IR) photometry and the optical discovery observations of the ``impostor" supernova (SN) 2010da in NGC~300 using new and archival \textit{Spitzer} \textit{Space Telescope} images and ground-based observatories. The mid-infrared counterpart of SN~2010da was detected as SPIRITS~14bme in the SPitzer InfraRed Intensive Transient Survey (SPIRITS), an ongoing systematic search for IR transients. Before erupting on May 24, 2010, the SN~2010da progenitor exhibited a constant mid-IR flux at 3.6 and only a slight $\sim10\%$ decrease at 4.5 $\mu$m between Nov 2003 and Dec 2007. A sharp increase in the 3.6 $\mu$m flux followed by a rapid decrease measured $\sim150$ d before and $\sim80$ d after the initial outburst, respectively, reveal a mid-IR counterpart to the coincident optical and high luminosity X-ray outbursts. At late times after the outburst ($\sim2000$ d), the 3.6 and 4.5 $\mu$m emission increased to over a factor of 2 times the progenitor flux and is currently observed (as of Feb 2016) to be fading, but still above the progenitor flux. We attribute the re-brightening mid-IR emission to continued dust production and increasing luminosity of the surviving system associated with SN~2010da. We analyze the evolution of the dust temperature ($T_\mathrm{d}\sim700 - 1000$ K), mass ($M_\mathrm{d}\sim0.5-3.8\times10^{-7}$ $\mathrm{M}_\odot$), luminosity ($L_\mathrm{IR}\sim1.3-3.5\times10^4$ $\mathrm{L}_\odot$) and the equilibrium temperature radius ($R_\mathrm{eq}\sim6.4-12.2$ AU) in order to resolve the nature of SN~2010da. We address the leading interpretation of SN~2010da as an eruption from a luminous blue variable (LBV) high-mass X-ray binary (HMXB) system. We propose that SN~2010da is instead a supergiant (sg)B[e]-HMXB based on similar luminosities and dust masses exhibited by two other known sgB[e]-HMXB systems. Additionally, the SN~2010da progenitor occupies a similar region on a mid-IR color-magnitude diagram (CMD) with known sgB[e] stars in the Large Magellanic Cloud. The lower limit estimated for the orbital eccentricity of the sgB[e]-HMXB ($e>0.82$) from X-ray luminosity measurements is high compared to known sgHMXBs and supports the claim that SN~2010da may be associated with a newly formed HMXB system.




\end{abstract}

\maketitle

\section{Introduction}

\begin{figure}[t]
	\centerline{\includegraphics[scale=.56]{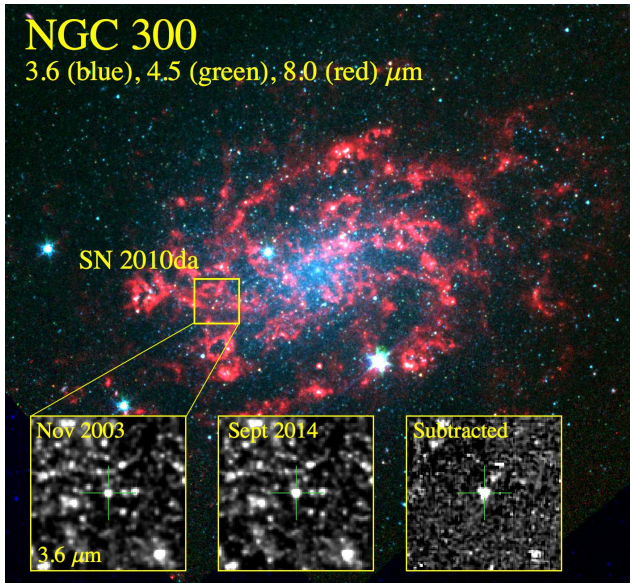}}
	\caption{\textit{Spitzer}/IRAC false color image of NGC~300 overlaid with the channel 1 (3.6 $\mu$m) zoom on SN~2010da/SPIRITS~14bme taken before and after the initial optical outburst. The ``Subtracted" inset shows the difference between the post-outburst and pre-outburst images.}
	\label{fig:SN2010da}
\end{figure}

The transient event designated supernova (SN) 2010da was discovered as an optical transient (OT) in the nearby galaxy NGC~300 (Fig.~\ref{fig:SN2010da}) on 2010 May 23 by Monard (2010). The initial classification as a SN, however, was immediately contradicted by its low outburst luminosity ($M_V\sim-10.3$ at maximum; Khan et al. 2010). Mid-infrared (IR) photometry from archival \textit{Spitzer} imaging data revealed that the progenitor of SN~2010da heavily obscured and shared a similar position in a mid-IR color-magnitude diagram (CMD) as luminous blue variable (LBV) candidates (Massey et al. 2007; Thompson et al. 2009; Khan et al. 2010). Follow-up optical spectroscopy of SN~2010da supported the claim that it was not a true SN, but instead resembled an LBV-like outburst (Elias-Rosa 2010; Chornock \& Berger 2010). Extreme outbursts from LBVs are typically associated with a class of SN impostors that are non-terminal stellar explosions with lower peak luminosities and ejecta velocities than typical Type~II SNe (Smith et al. 2011; van Dyk \& Matheson 2012). 

The LBV eruption hypothesis for SN~2010da, however, is confounded by low progenitor luminosities ($\sim10^4$ $\mathrm{L}_\odot$; Prieto et al. 2010; this work) and the detection of an X-ray point source from \textit{Swift} within hours of the optical detection of SN~2010da (Immler et al. 2010) that exhibited a 0.3-10 keV luminosity of $\sim5\times10^{38}$ erg $\mathrm{s}^{-1}$. Studies of the subsequent X-ray emission by Binder et al. (2016) revealed a recurring, less luminous ($L_x\sim4\times10^{37}$ erg $\mathrm{s}^{-1}$; 0.35-8 keV) outburst with \textit{Chandra}. These X-ray luminosities are several orders of magnitude greater than the highest observed X-ray luminosities from an LBV outburst ($\sim10^{34}$ erg $\mathrm{s}^{-1}$, Naz\'{e} et al. 2012) and a massive colliding-wind binaries ($\sim10^{35}$ erg $\mathrm{s}^{-1}$,e.g. Guerrero \& Chu 2008). In order to reconcile the X-ray observations with the mid-IR and optical studies that implicated an LBV-like eruption, Binder et al. (2016) suggest that SN~2010da is a high-mass X-ray binary (HMXB) composed of a neutron star and an LBV-like companion. Their claim is substantiated by the presence of a young cluster in the vicinity of SN~2010da estimated to be $\lesssim5$ Myr-old based on analysis of its color-magnitude diagram (CMD). Binder et al. (2016) propose the interesting hypothesis that the initial outburst of SN~2010da marked the onset of the HMXB phase based on the recurring X-ray bursts and the presumed quiescence of the progenitor, where $\mathrm{L}_\mathrm{X}<(3-9)\times10^{36}$  erg $\mathrm{s}^{-1}$ at 4 epochs.


HMXBs are accretion-powered binary systems with a neutron star or black hole orbiting an O or B-type companion and are subdivided into two subclasses depending on the evolutionary status of the optical companion (Liu et al. 2005; Reig 2011). The most common types of HMXBs ($\sim60\%$; Liu et al. 2006) are known as Be/X-ray binaries (BeXBs) and host a rapidly-rotating, non-supergiant B-type star with low-ionization emission lines attributed to an equatorial ``excretion" disk of material ejected from the star. Variable X-ray emission from BeXBs is associated with the interactions between the circumstellar disk around the Be star and the wide, eccentric orbits of the compact object companion. The less commonly detected subclass of HMXBs host O or B-type supergiant optical companions (sgHMXBs), where the X-ray emission is powered by accretion onto the compact companion from Roche-lobe overflow and/or stellar wind capture. Wind-fed sgHMXBs exhibit persistent X-ray luminosities of $\sim10^{35-36}$ erg $\mathrm{s}^{-1}$ while the systems that accrete from Roche-lobe overflow can result in higher luminosity outbursts ($\sim10^{38}$ erg $\mathrm{s}^{-1}$; e.g. Chaty et al. 2008). 

Recent hard X-ray observations ($\gtrsim15$ keV) less affected by extinction have revealed an emerging population of highly obscured ($A_V\gtrsim15$) sgHMXBs (e.g. Filliatre \& Chaty 2004; Chaty et al. 2008; Coleiro et al. 2013). The most intriguing are the obscured systems that host supergiant (sg)B[e]-stars, supergiant counterparts of Be stars with forbidden emission lines. SgB[e]-stars appear spectroscopically similar to LBVs and occupy similar regions of the HR diagram (e.g. Smith 2014); however, they do not exhibit extreme variability in photospheric temperatures or giant eruptions associated with LBVs. Their similar spectral properties may suggest there is an evolutionary link between the two phases. Studies of such a system would be beneficial for understanding the dusty environment of highly obscured sgHMXBs. Resolving the nature of SN~2010da is therefore one of the main purposes of this paper. 

In this paper, we present optical and mid-IR observations of SN~2010da and discuss the evolution of its optical and mid-IR properties (e.g. temperature and luminosity) inferred from the light curves. We address the nature of SN~2010da by proposing various interpretations (e.g. LBV eruption, merger product, HMXB) and assessing their validity given the energetics and dust properties we derive from observations. We adopt a distance of 2.0 Mpc ($\mu=26.5$) towards NGC~300 (Dalcanton et al. 2009).



\section{Observations}

\subsection{Optical Discovery and Follow-up Observations}

\subsubsection{Discovery from Bronberg Observatory}

\begin{figure}[t]
	\centerline{\includegraphics[scale=0.42]{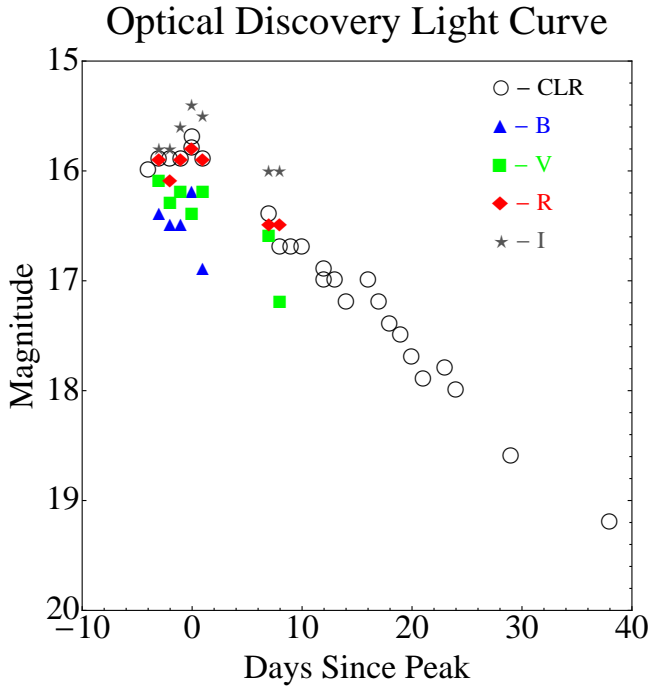}}
	\caption{Multi-band Optical discovery light curve of SN~2010da taken from Bronberg Observatory. The peak at $t=0$ corresponds to the time of peak optical brightness (JD = 2455343.66).}
	\label{fig:SN2010BMLC}
\end{figure}

The outburst of a new optical transient (OT) in NGC~300 was discovered by Monard (2010)
with his 0.3-m telescope at the Bronberg Observatory and an
unfiltered ST7-XME CCD camera. The initial detection was made on 2010 May 23.16,
during a morning session of SN searching in nearby galaxies. A confirmation
image was obtained on 2010 May 24.14 with the same instrumentation. These were
the first Bronberg observations of NGC~300 of the season, following its solar
conjunction, showing that the eruption was already underway. The discovery light curve is shown
in Fig.~\ref{fig:SN2010BMLC} and the data are provided in Tab.~\ref{tab:BMonard}.

\subsubsection{Follow-up with SMARTS}

On 2010 May 25 we began photometric monitoring of this event with the ANDICAM
CCD camera on the SMARTS Consortium\footnote{SMARTS is the Small \& Moderate
Aperture Research Telescope System; {\tt http://www.astro.yale.edu/smarts}}
1.3-m telescope at Cerro Tololo Inter-American Observatory (CTIO). This
monitoring continued until 2010 September 5, after which the OT became too faint
for useful observations.  A re-brightening of the object
was detected on 2011 October 21.16 by Chornock et al.\ (2011). We resumed the
1.3-m coverage on 2011 November 12, and continued until 2011 December 20,
after which the OT was again too faint for further observations.

The SMARTS monitoring was carried out by service personnel at CTIO, using {\it
BVRI\/} filters in the Johnson-Kron-Cousins system. The ANDICAM frames were
bias-subtracted and flat-fielded in the SMARTS pipeline at Yale University. We
then determined differential magnitudes between the OT and a nearby comparison
star, using standard aperture-photometry tasks in IRAF\footnote{IRAF is
distributed by the National Optical Astronomy Observatory, which is operated by
the Association of Universities for Research in Astronomy (AURA) under a
cooperative agreement with the National Science Foundation.}. Absolute
photometry of the comparison star (J2000: 00:55:06.47, -37:42:58.2) was
determined from aperture photometry on 11 sets of {\it BVRI\/} frames obtained
on four photometric nights, calibrated to standard stars of Landolt (1992). This
calibration yielded $V=13.710$, $B-V=0.596$, $V-R=0.346$, and $V-I=0.675$, with
mean errors of about $\pm$0.003--0.007~mag.

The resulting calibrated magnitudes of the OT from the SMARTS monitoring are
listed in Table~\ref{tab:FluxO} and are shown in Figs.~\ref{fig:SN2010daLC}B, \ref{fig:BVRILC}A and \ref{fig:BVRILC}B. The tabulated uncertainties are based only on photon
statistics; the systematic errors due to the zero-point uncertainty and
instrumental effects are likely of order $\pm$0.01--0.02~mag. In addition, the
OT lies in a small cluster (Binder et al.\ 2016), only marginally resolved in
the 1.3-m images, which contaminates the photometry at the OT's faintest level;
we have not attempted to correct for this effect.

\subsubsection{Follow-up with Swope}

The $g$,$r$, and $i$-band follow-up images were taken between 2014 May 17 to 2015 November 13 with the 1m-Swope telescope, at Las
Campanas Observatory, using a recently installed e2v CCD, 4K$\times$4K pixels array of 0.435"/pix.
The observations, the standard data reduction procedure (bias subtraction,
flat fielding and linearity correction) and the photometry calibration was performed
by Carnegie Supernova Project collaborators in similar way as detailed in Hamuy et al. (2006). 
The field was calibrated in 2 photometric nights: 2014 August 1 and  2014 December 20.
The PSF photometry of SN~2010da were converted from the Natural system of Swope (Contreras et al. 2010) to AB magnitudes based on the transformations provided by Eq.~A5-A7 in Stritzinger et al. (2011). The following color terms (CT) were adopted in these transformations: $\mathrm{CT}_g = -0.010$, $\mathrm{CT}_r = -0.014$, and $\mathrm{CT}_i = +0.018$. Since the transformed AB magnitudes of SN~2010da were consistent within $<0.01$ mag of the Natural system photometry, the Natural magnitudes were treated as AB magnitudes when ultimately converting to the Vega system presented in Tab.~\ref{tab:FluxO}. The $gri$ photometry is shown in Figs.~\ref{fig:SN2010daLC}B and \ref{fig:BVRILC}C.

\subsection{Mid-Infrared Observations with \textit{Spitzer}}

The SPitzer InfraRed Intensive Transient Survey (SPIRITS; Kasliwal et al. 2016, in prep.) targets 194 nearby galaxies within 20 Mpc to a depth of 20 mags on the Vega system. The observations are performed in the 3.6 $\mu$m and 4.5 $\mu$m bands of the InfraRed Array Camera (IRAC, Fazio et al. 2004) on board the warm \textit{Spitzer Space Telescope} (Werner et al. 2004; Gehrz et al. 2007). 

The mid-IR counterpart to the SN~2010da, designated SPIRITS~14bme, was discovered by SPIRITS in \textit{Spitzer}/IRAC imaging with channel 1 ($3.6$ $\mu$m) and channel 2 (4.5 $\mu$m) taken on 2014 September 5. SPIRITS~14bme is located at a right ascension and declination of 00:55:04.87 and -37:41:43.8, within $<0.2''$ of the optical counterpart (Monard 2010).  The progenitor is detected in archival IRAC taken with channels 1 - 4 (3.6, 4.5, 5.8, and 8.0 $\mu$m) on 2003 November 21, 2007 December 28.88, and 2007 December 28.94 before the warm mission. All of the available channel 1 and 2 imaging data of SN~2010da/SPIRITS~14bme from the \textit{Spitzer} Heritage Archive taken between 2003 November and 2016 March were incorporated in this paper.

To obtain magnitudes, forced aperture photometry was performed on the Post Basic Calibrated Data. The flux was summed in a 4-pixel aperture centered at the SPIRITS-determined coordinates of SPIRITS~14bme. Sky background was measured within an annulus from 8 to 16 pixels surrounding each source and subtracted from the total flux. Finally, fluxes were converted to magnitudes using the Warm \textit{Spitzer}/IRAC zero points of 18.8024 (channel 1) and 18.3174 (channel 2), along with aperture corrections of 1.21 and 1.22, respectively, as specified by the IRAC instrument handbook. The channel 1 and 2 fluxes of all available IRAC imaging of SN~2010da/SPIRITS~14bme are provided (in $\mu$Jy) in Tab.~\ref{tab:Flux}, and the photometry is shown in Fig.~\ref{fig:SN2010daLC}A. The channel 3 and 4 fluxes of the SN~2010da progenitor taken on November 21, 2003 are $87.93\pm28.30$ $\mu$Jy and $54.89\pm25.00$ $\mu$Jy, respectively. The progenitor was also observed with the Multiband-Imaging Photometer for \textit{Spitzer} (MIPS) on November 21, 2003 from which an upper limit of $\lesssim136$ Jy at 24 $\mu$m is derived.

In the following sections of this paper, we refer to SPIRITS~14bme as SN~2010da or its mid-IR counterpart.


\subsection{Near-Infrared Observations}

The goal of the SPIRITS campaign at the University of Minnesota's Mount Lemmon Observing Facility (MLOF) in Arizona is two-fold. First, transients of interest from the space-based SPIRITS observations are followed up in the J ($1.25$ $\mu$m), H ($1.65$ $\mu$m) and $\mathrm{K}_\mathrm{s}$ ($2.15$ $\mu$m) IR bands using a 2MASS imaging camera (Milligan et al. 1996, Skrutskie et al. 2006) mounted on the 60'' telescope at MLOF. This provides additional photometric data on the temporal light curves for these objects. Secondly, a general survey of potential SPIRITS targets are imaged on the 2MASS detector, providing potential information on the progenitors of future SPIRITS objects, and possibly even the discovery of new objects should they be luminous enough for ground-based observations. Sixty minutes of short exposure images of the targeted SPIRITS objects are obtained with the 2MASS camera, along with 60 minutes of images of surrounding sky. Using these sky images to remove background radiation, the science images are stacked, co-added, and analyzed to obtain photometric data. Limiting magnitudes with respect to the background are typically $\sim16-17$ mag for H and $\mathrm{K}_\mathrm{s}$ bands, and $\sim17-19$ mag for the J band.

J, H, and $\mathrm{K}_\mathrm{s}$ images of NGC~300 were taken at MLOF on 2014 November 23 and 2015 November 22. On 2014 November 23, the J, H, and $\mathrm{K}_\mathrm{s}$ magnitudes for SN~2010da were $17.5\pm0.2$, $16.6\pm0.3$, and $16.1\pm0.4$, respectively. On 2015 November 22, the J, H, and $\mathrm{K}_\mathrm{s}$ magnitudes were $18.0\pm0.3$, $17.3\pm0.4$, and $16.7\pm0.5$, respectively. 


Near-IR J and $\mathrm{K}_\mathrm{s}$ images of NGC~300 were also obtained on 2014 December 24 with the FourStar infrared camera (Persson et al. 2013) on the Magellan Baade Telescope at Las Campanas Observatory in Chile. The measured J and $\mathrm{K}_\mathrm{s}$ magnitudes for SN~2010da from FourStar were $18.1\pm0.3$ and $16.5\pm0.2$, respectively.

\section{Results and Analysis}

\begin{figure}[t]
	\centerline{\includegraphics[scale=0.37]{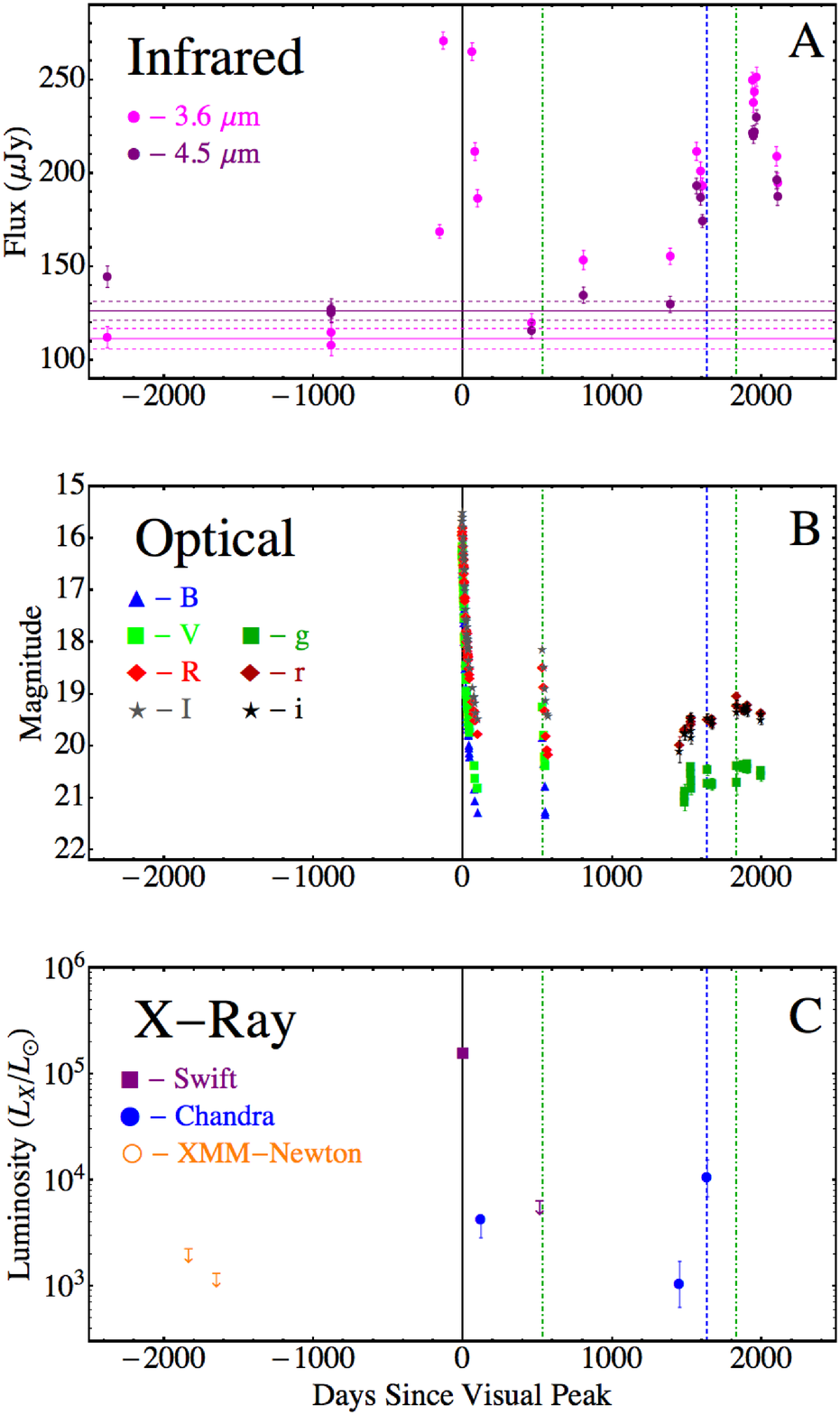}}
	\caption{Mid-IR (A), optical (B), and X-ray (C) light curve of SN~2010da. X-ray data are from Binder et al. (2016 and ref. therein). Green dot-dashed lines and blue dashed lines correspond to observed peaks in optical and X-ray emission, respectively.}
	\label{fig:SN2010daLC}
\end{figure}

\subsection{SN 2010da Light Curves} 

\subsubsection{Mid-Infrared Photometry} 

The full mid-IR light curve of SN 2010da taken by \textit{Spitzer}/IRAC before, during, and after the optical and X-ray outburst is shown in Fig.~\ref{fig:SN2010daLC}A, where $t_d=0$ corresponds to the day of the optical peak (JD = 2455343.66). The optical light curve (Fig.~\ref{fig:SN2010daLC}B) and X-ray observations (Fig.~\ref{fig:SN2010daLC}C) from Binder et al. (2016 and ref therein) are plotted on the same time axis as the mid-IR light curve. On day -156, the 3.6 $\mu$m emission increased by $\sim50\%$ of the progenitor flux as measured from day -2379 and -880 observations. This flux increase is significant given that the source has remained constant to within a few percent at 3.6 and within $\sim10\%$ at 4.5 $\mu$m between days -2379 and -880. Within 23 days of the initial brightening on day -156, the 3.6 $\mu$m flux increased to its highest measured mid-IR value of 0.27 mJy, a factor of $\sim2.5$ times the progenitor flux. At day 60 after the initial optical outburst, the source exhibited a similar 3.6 $\mu$m flux as day -133. The sharp increase and immediate decrease in brightness at 3.6 $\mu$m indicates that an outburst in the mid-IR occurred between days -133 and 81, which is consistent with the timing of the X-ray and optical emission peaks.

By day 460, the 3.6 and 4.5 $\mu$m flux decreased to values slightly higher and lower than the progenitor flux at respective wavelengths. The following measurement made on day 806 indicated a gradual re-brightening in the mid-IR with similar fluxes observed on day 1386. Between days 1386 and 1602, there is evidence of rapid re-brightening given the sharp increase in flux from day 1386 to 1562 followed by a steep decrease. At day 1936, the mid-IR flux increased by $\sim25\%$ from day 1602 and decreased slightly in the subsequent observation on day 1943. This is suggestive of another re-brightening event in the mid-IR between days 1602 and 1936 and, as in the case of the initial outburst, appears to coincide with the re-brightening X-ray and optical emission (blue dashed and green dot dashed lines in Fig.~\ref{fig:SN2010daLC}, respectively). The following two observations on days 1951 and 1965 show a rapid increase to the peak measured brightness at late-times, almost a factor of $\sim2$ greater than the progenitor flux. The most recent two mid-IR observations made on days 2097 and 2104 show a sharp decrease in flux $\sim10\%$ lower than on day 1965, which suggests that another peak in the mid-IR emission occurred between days 1965 and 2097.



\subsubsection{Optical Photometry} 

\begin{figure*}[t]
	\centerline{\includegraphics[scale=0.48]{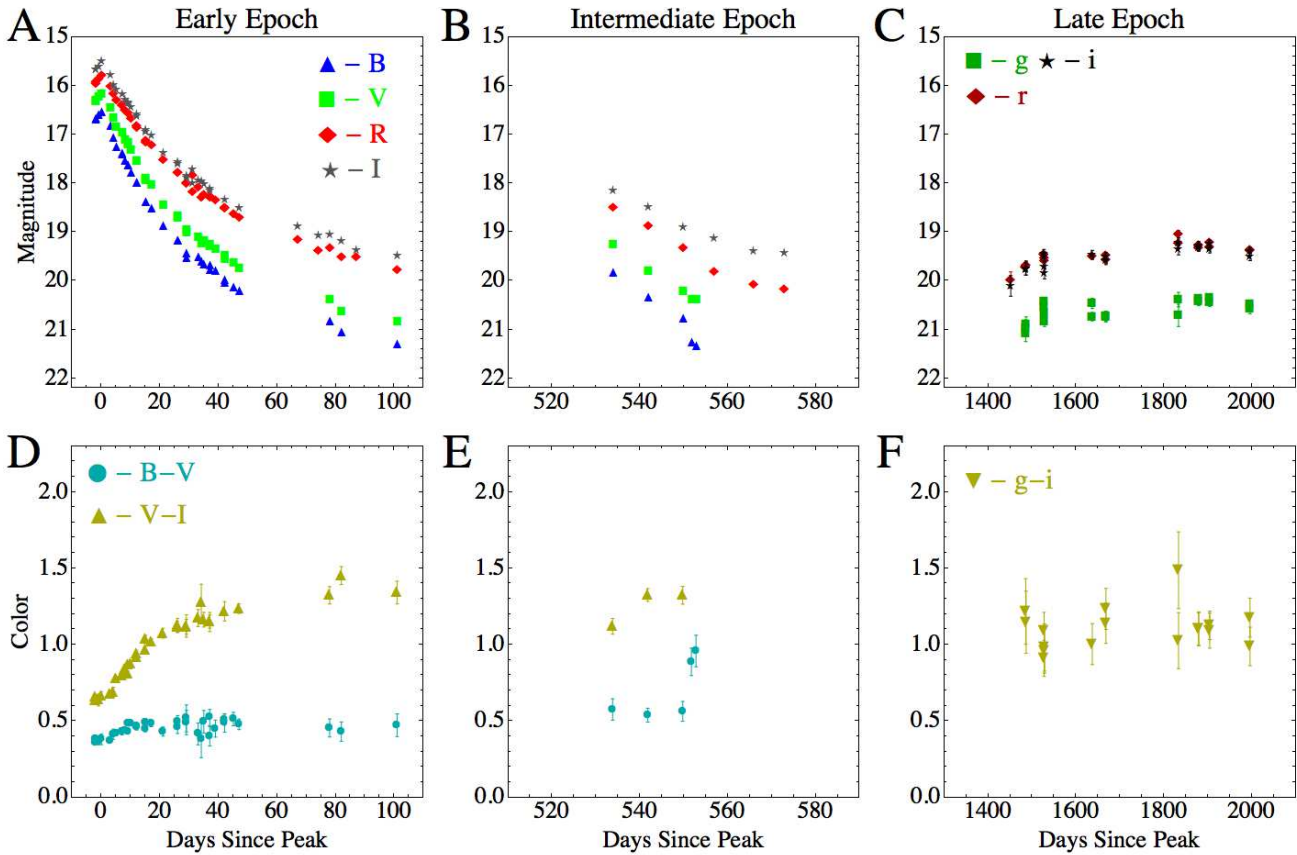}}
	\caption{Early (A), intermediate (B), and late (C) epoch optical light curves and color evolution (D) of SN~2010da. The inset shows a zoom of the early-time colors. Early and intermediate $BVRI$ light curves were taken from SMARTS-1.3m telescope. The late time $gri$ light curve was taken from the 1m Swope telescope.}
	\label{fig:BVRILC}
\end{figure*}

Detailed $BVRI$ light curves of SN~2010da taken at ``early" (day $\sim0-100$) and ``intermediate" (day $\sim500$) epochs and the $gri$ light curve taken at a ``late" (day $\sim1300-2000$) epoch are shown in Fig.~\ref{fig:BVRILC}A, B, and C, respectively. Prior to its discovery as an optical transient (Monard 2010), archival optical imaging of NGC~300 from Magellan/Megacam reveal non-detections of r and i-band counterparts to a deep limit of 24 AB mag (Berger \& Chornock 2010). The early epoch observations capture a peak in the optical emission occurring 4.2 d after its discovery followed by a rapid decline in brightness by a factor of $\sim20$ from the peak flux within 50 d, and a factor of $\sim60$ within 100 d. Given the peak V-band flux of $16.17\pm 0.02$ mag and a distance modulus of $\mu=26.5$ (Dalcanton et al. 2009), the absolute visual magnitude of the optical peak is $M_V=-10.3$, consistent with the value reported from the initial discovery (Khan et al. 2010; Monard 2010). 


The intermediate epoch $BVRI$ light curve (Fig.~\ref{fig:BVRILC}B) reveals a re-brightening of the optical counterpart to $V=19.27\pm0.04$ mag, which is $\sim3.0$ mag fainter than the peak brightness but $\sim1.5$ mag brighter than the final $V$ magnitude measured in the early epoch. The timing and brightness of this re-brightening is consistent with the 6.5-m Magellan/Clay telescope optical observations of SN 2010da reported by Chornock, Czekala, \& Berger (2011). As in the case of the initial outburst, the intermediate outburst exhibits a steep decline in brightness in the following $\sim50$ d. 

The late-time $gri$ light curve (Fig.~\ref{fig:BVRILC}C) exhibits a gradual re-brightening between days 1530 and 1833 by $\sim0.5$ mag in all three bands. Only the $r$ band, however, exceeds the lowest measured brightness in the earlier epochs assuming $r\approx R$, $g\approx V$, and $i\approx I$. The late epoch optical re-brightening is consistent with the timing of the re-brightening X-ray emission observed on day 1635. In the $\sim70$ days following the apparent optical peak at day 1833, the optical brightness does not vary significantly. However, the final $gri$ measurements made on day 1996, which is $\sim30$ d after the final mid-IR observation, show a $\sim0.2$ mag decrease from the peak brightness.

The evolution of the optical colors ($B-V$, $V-I$, and $g-i$) from the outbursts at early, intermediate, and late epochs is shown in Fig.~\ref{fig:BVRILC}D, E, and F, respectively. In the early epoch, the $V-I$ color becomes significantly redder within $\sim20$ d of the initial outburst, increasing from $V-I\approx0.6$ to 1.0. The $B-V$ color does not exhibit the same evolution and only increases from $B-V\approx0.38$ at the outburst to 0.45 after 20 days. The optical colors measured at the peak of the intermediate epoch are consistent with the colors in the early epoch after day 40; however, there are less than 10 observations between days 40 and 530. At the final $B$ and $V$-band measurements of the decreasing brightness after the outburst, the $B-V$ color shows a rapid red-ward change from $B-V\approx0.48$ to 0.92 within a short 10-day timescale. Late time $g-i$ colors are consistent with the $V-I$ colors at intermediate times ($g-i\sim1.1$). 





\begin{figure*}[t]
	\centerline{\includegraphics[scale=0.5]{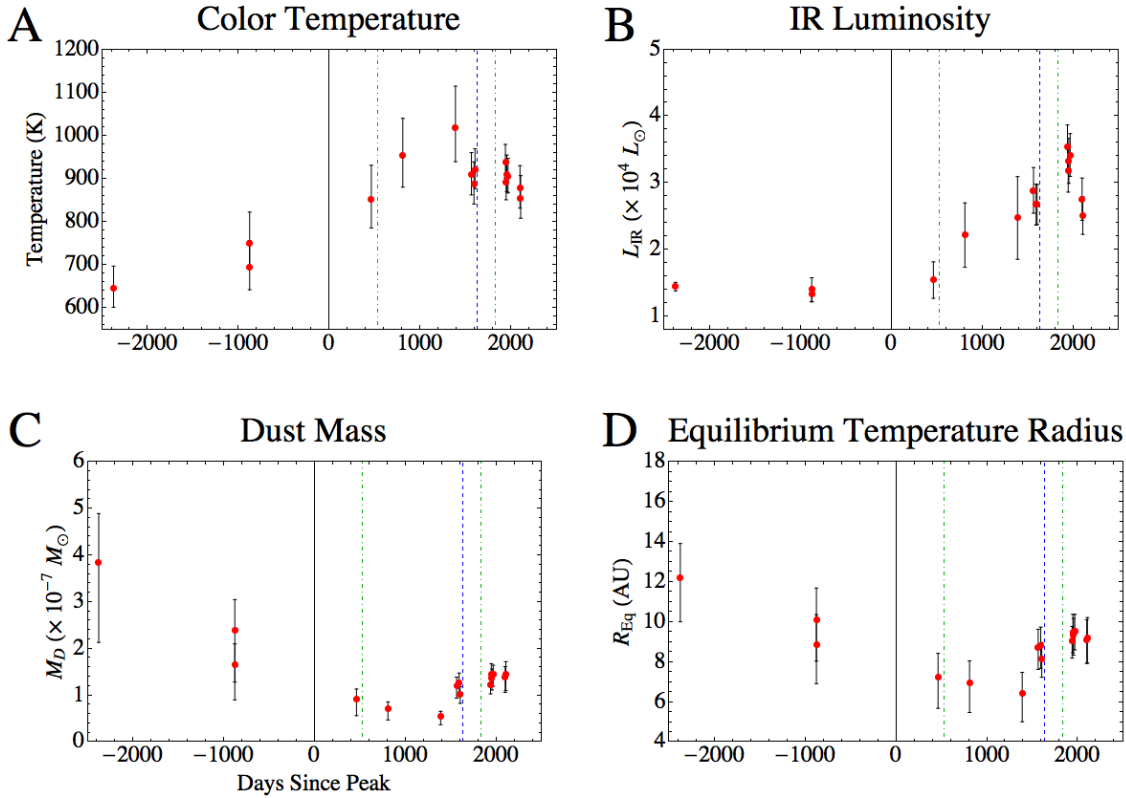}}
	\caption{Evolution of the color temperature (A), IR luminosity (B), dust mass (C), and equilibrium temperature radius (D) derived from the mid-IR emission. Green dot-dashed lines and blue dashed lines correspond to observed peaks in optical and X-ray emission, respectively.}
	\label{fig:TLM}
\end{figure*}

\subsection{Properties of the SN~2010da Mid-IR and Optical Counterpart}

\subsubsection{Dust Temperature, IR Luminosity, Mass, and Equilibrium Temperature Radius}
\label{sec:dprop}

The detection of the progenitor in the mid-IR and non-detection at optical wavelengths are consistent with a central source that was enshrouded by dense and dusty circumstellar material. This mid-IR emission likely originates from hot ($T_D\gtrsim500$ K) circumstellar material in the vicinity of SN 2010da and its progenitor and can therefore be used to trace the properties of the emitting dust. The temporal evolution of the dust temperature ($T_D$), total IR luminosity ($L_\mathrm{IR}$), and mass ($M_D$) and the equilibrium temperature radius ($R_\mathrm{Eq}$) are shown in Fig.~\ref{fig:TLM}A-D, respectively, and provided in Tab.~\ref{tab:Flux}.


Dust temperatures were derived by fitting the 3.6 and 4.5 $\mu$m fluxes to a single-temperature modified blackbody, a Planck function multiplied by a grain emissivity model assuming a composition of carbonaceous $a=0.1$ $\mu$m-sized grains (Compi\`{e}gne et al. 2010). The progenitor dust exhibited temperatures of $650-750$ K. At the first detection made with at both 3.6 and 4.5 $\mu$m after the outburst (day 460), the dust temperature increased to $850$ K and showed an increasing trend up until it peaked at day 1386 to $1020$ K. The following measurement taken 175 days after the temperature peak revealed that the temperature quickly decreased to $910$ K. No significant deviations from $\sim900$ K were measured, but the most recent observation on day 2104 indicate a possible decrease to $\sim850$ K.

The total infrared luminosity of the emitting dust was estimated by integrating over the temperature-fitted modified blackbody and assuming a distance of 2.0 Mpc towards NGC~300 (Dalcanton et al. 2009). The evolution of $L_\mathrm{IR}$ closely resembles the mid-IR light curve with a gradual post-outburst increase followed by a decrease measured at day 2097. Prior to the eruption, the progenitor exhibited an IR luminosity of $L_\mathrm{IR}\approx1.4\times10^4$ $\mathrm{L}_\odot$, consistent with the mid-IR progenitor luminosity determined by Prieto et al. (2010). On day 1936, where measured IR luminosity peaked, $L_\mathrm{IR}$ increased by over a factor of two from from the progenitor levels to $3.5\times10^4$ $\mathrm{L}_\odot$. The final measurements made on days 2097 and 2104 show a decrease from the peak luminosity to $\sim2.5\times10^4$ $\mathrm{L}_\odot$.

Dust masses were derived from the mid-IR fluxes and temperature-fitted modified blackbodies as indicated in Eq.~\ref{eq:Mass}:

\beq
M_D=\frac{(4/3)\,a\,\rho_b\,F_\lambda\,d^2}{Q_C(\lambda,a )\,B_\lambda(T_D)},
\label{eq:Mass}
\eeq

\noindent
where $a$ is the adopted grain size, $\rho_b$ is the bulk density of the dust grains, $F_\lambda$ is the measured flux, $d$ is the distance to the source, $Q_C(\lambda,a)$ is the grain emissivity model for carbonaceous grains, and $B_\lambda(T_D)$ is the Planck function. A bulk density of $\rho_b=2.2$ $\mathrm{gm}$ $\mathrm{cm}^{-3}$ (e.g. Draine \& Li 2007) is assumed for the emitting carbonaceous grains. The estimated dust mass of the progenitor ($M_D\sim2-4\times10^{-7}$ $\mathrm{M}_\odot$) decreased by over a factor of $\sim2$ at the first 3.6 and 4.5 $\mu$m measurement made after the outburst at day 460. In the following $\sim 930$ days, the dust mass gradually decreased to a minimum of $0.54\times10^{-7}$ $\mathrm{M}_\odot$. Between days 1386 and 1562, the dust mass increased significantly by $0.7\times10^{-7}$ $\mathrm{M}_\odot$, which implies dust was formed at a rate of $\mathrm{\dot{M}}_D\approx1.4\times10^{-7}$ $\mathrm{M}_\odot$ $\mathrm{yr}^{-1}$. The timing of this increased dust production is consistent with the late time re-brightening in the X-ray, optical, and mid-IR as well as the decrease in dust temperature. The final mid-IR measurements taken around days $\sim2000$ reveal a slight ($\sim20\%$) increase in dust mass from day 1602.

\begin{figure}[t]
	\centerline{\includegraphics[scale=0.65]{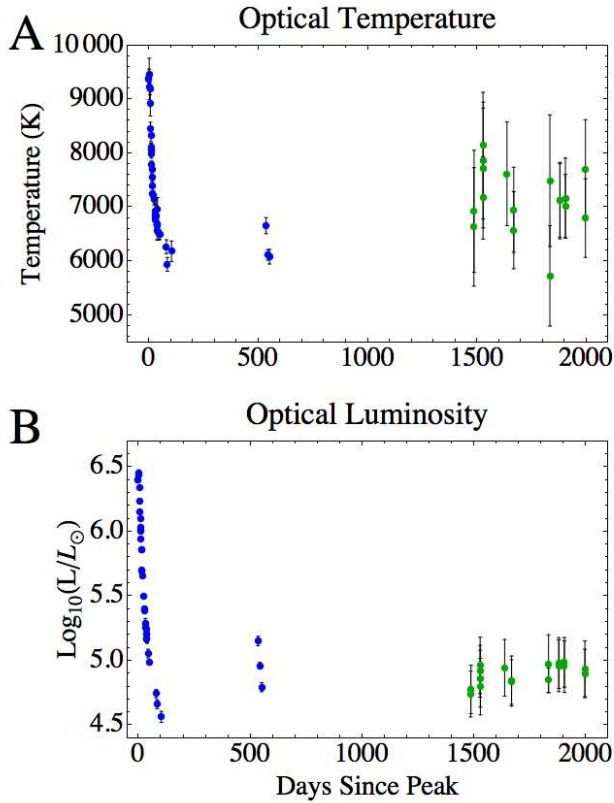}}
	\caption{Temperature (A) and luminosity (B) of optical counterpart determined from \textit{BVI} (blue points) and \textit{gi} (green points) photometry. $R$ and $r$ band photometry were neglected in the temperature and luminosity estimates due to possible contamination from H$\alpha$ line emission.}
	\label{fig:BVRILT}
\end{figure}

The equilibrium temperature radius, $R_\mathrm{Eq}$, is determined from the distance where the absorbed power from the radiation field of the heating source is equal to the power emitted by the dust grain of temperature $T_D$. The equilibrium temperature radius provides an estimate of the physical size of the hot, circumstellar dust and is given by

\beq
R_\mathrm{Eq}\approx\left(\frac{Q_\mathrm{abs}}{Q_\mathrm{e}}\frac{L_\mathrm{IR}}{16\pi\sigma T^4_D}\right)^{1/2},
\label{eq:temp}
\eeq
\noindent
where $Q_\mathrm{abs}$ and $Q_\mathrm{e}$ are the grain absorption and emission Planck mean cross sections (see Gilman 1974), respectively, and $\sigma$ is the Stefan-Boltzmann constant. A typical value of $\left(Q_\mathrm{abs}/Q_\mathrm{e}\right)\approx0.3$ (e.g. Smith et al. 2016) is assumed for the calculation. Note that if the heating of the mid-IR emitting dust is instead dominated locally by collisions with hot electrons in a post-shock environment (e.g. Dwek 1987, Dwek et al. 2008), $R_\mathrm{Eq}$ will not reflect the physical size of the system. However, it is unlikely that the mid-IR emitting dust is dominantly heated by a shock driven from the initial outburst since the observed mass of an illuminated shell of shock-heated dust should increase rapidly as a function of time as it propagates outward radially ($M_\mathrm{D}\propto t^2$; Fox et al. 2011). This is inconsistent with the ``step-function"-like evolution of the dust mass.

Before the outburst, the progenitor exhibit an equilibrium temperature radius of $\sim10-12$ AU. After the outburst on day 460, the radius decreased from the progenitor size to $\sim7$ AU and did not change significantly in the following measurements made on days 806 and 1386. At late-times coincident with the multi-wavelength outbursts and dust mass increase, the radius grew to $\sim9$ AU and did not deviate significantly throughout the final observations. 


\subsubsection{Optical Counterpart Temperature and Luminosity}

The evolution of the temperature and luminosity of the dereddened SN 2010da optical counterpart is shown in Fig.~\ref{fig:BVRILT}A and B, respectively, and are provided in Tab.~\ref{tab:FluxO}. Optical photometry was corrected for foreground extinction ($A_V=0.26$) and the differential extinction internal to NGC~300 along the line of sight towards SN~2010da ($A_V\sim0.4$; Binder et al. 2016).

Temperatures were derived by fitting a single-temperature blackbody to the $BVI$ and $gi$ photometry, where the $R$ and $r$-bands are purposely ignored due to potential contamination from prominent $H\alpha$ line emission detected from the outburst (Chornock \& Berger 2010). Temperatures peak at $\sim9500$ K for several days around the initial optical peak and decrease rapidly to $\sim6500$ K in the 40 days following the peak. At the intermediate epoch, the temperature peaks at $\sim6700$ K on day 534 and decreases to $\sim6200$ K within 8 days. Temperatures derived in the late epoch have large error bars due to the faintness of the optical counterpart; however, they exhibit a range of temperature $\sim6000-8000$ K consistent with values estimated during the previous epochs. 

Luminosities of the optical counterpart to SN 2010da were estimated by integrating over the temperature-fitted blackbodies. The highest measured luminosity is $3\times10^6$ $\mathrm{L}_\odot$ and is consistent with the initial optical brightness peak. After the peak, the luminosity rapidly decreased by over two orders of magnitude to $\sim4\times10^4$ $\mathrm{L}_\odot$ at the end of the early epoch ($t_d= 101$ d). At the intermediate time re-brightening on day $\sim530$, the luminosity peaked at $\sim1.4\times10^5$ $\mathrm{L}_\odot$ and decreased by a factor of $\sim2$ in the following $40$ days. In the late epoch, the optical counterpart exhibited luminosities ranging from $\sim5-10\times10^4$ $\mathrm{L}_\odot$, which is consistent with the luminosity minimum of the intermediate epoch. 


\subsection{Progenitor and Post-Outburst DUSTY SED Models}

\begin{figure}[t]
	\centerline{\includegraphics[scale=0.63]{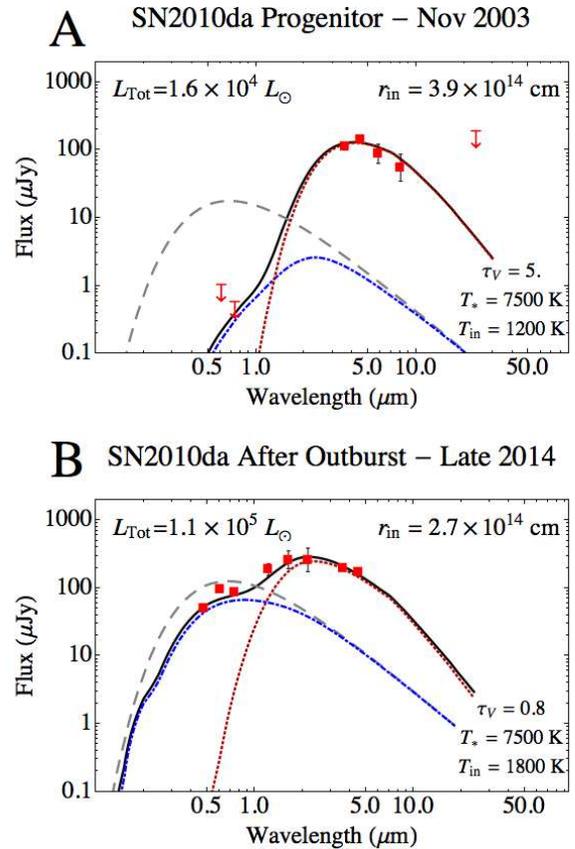}}
	\caption{DUSTY radiative transfer models of the SN~2010da progenitor (A) and after the outburst (B). The gray dashed line represents the unabsorbed spectrum from the heating source. The blue dot-dashed and red-dot dashed lines correspond to emission from attenuated stellar heating source and dust emission, respectively. The black solid line indicates the total spectra outputted from the system.}
	\label{fig:Dmod}
\end{figure}

Spectral energy distributions (SEDs) from multi-wavelength photometry of SN 2010da and its progenitor can be modeled with the radiative transfer code DUSTY (Ivezic \& Elitzur 1997) to provide additional constraints the properties of the emitting dust and heating source before and after the outburst. DUSTY models are fitted to flux measurements taken on day -2379 and between days 1602-1641\footnote{$gri$ and $JHK_\mathrm{S}$ observations were taken on days 1636.7 and 1641.1, respectively, and the closest 3.6 and 4.5 $\mu$m observation was made on day 1602}, which are referred to as the progenitor and post-outburst models, respectively. For both models, dust is assumed to be in a spherical shell with radial density profile proportional to $r^{-2}$ and made up of amorphous carbon with an MRN (Mathis, Rumpl, \& Nordseick 1977) grain size distribution ($a_\mathrm{min}=0.005$ $\mu$m, $a_\mathrm{max}=0.25$ $\mu$m, and a grain-size power law proportional to $a^{-3.5}$). Based on the measured temperatures of the optical counterpart (Fig.~\ref{fig:BVRILT}B), a 7500 K blackbody is adopted as the central heating source for both models. The free parameters are the heating source luminosity ($L_*$), optical depth through the dust shell ($\tau_V$), shell thickness as a multiplicative factor, $Y$, of inner radius size, $r_\mathrm{in}$, and dust temperature at the inner radius ($T_{\mathrm{in}}$)\footnote{$r_\mathrm{in}$ corresponds to $T_{\mathrm{in}}$ in DUSTY so they are not independent variables.}. Following the analysis of Khan et al. (2015), the fitted parameters from DUSTY can be used to estimate the total mass of the shell, $M_\mathrm{e}$, as follows:

\beq
M_\mathrm{e}=\frac{4\pi\,r_\mathrm{in}^2\,\tau_V}{\kappa_V},
\label{eq:mass}
\eeq
\noindent
where $\kappa_V$ is the V-band opacity of the dust which is assumed to be $100$ $\mathrm{cm}^2$ $\mathrm{g}^{-1}$. Due to the unresolved nature of the emission from SN~2010da, there is a lack of morphological information which limits the utility of the DUSTY models. $Y$ and $r_{\mathrm{in}}$ are therefore not well constrained by the mid-IR fluxes: increasing $Y$ broadens the model SED as well as shifts the peak to longer wavelengths, which can be balanced by increasing $r_{\mathrm{in}}$. There is also a degeneracy between $r_{\mathrm{in}}$ and $\tau_V$ since enhanced near/mid-IR emission from hotter dust can be mitigated by increasing optical depth. The upper limit models on the parameters $r_{\mathrm{in}}$ and $M_\mathrm{e}$ are presented in Fig.~\ref{fig:Dmod}A and B with the values given in Tab.~\ref{tab:DUSTY}. 

Note that the true morphology of the dust in the system may not resemble the spherical geometry assumed for the DUSTY models and likely differs before and after the initial outburst. Additionally, due to the lack of far-IR photometry, the luminosity and optical depth derived by DUSTY assumes there is only a single distribution of hot dust traced by the mid-IR emission. This derived heating source luminosity, however, is well-defined since the observed optical to mid-IR flux measurements and limits provide robust constraints on the shape and amplitude of the warm dust SED. 

The progenitor model shown in Fig.~\ref{fig:Dmod}A indicates that the luminosity of the heating source is $L_*=1.6\times10^4$ $\mathrm{L}_\odot$, consistent with the integrated IR luminosity inferred from the analysis in Sec.~\ref{sec:dprop} (Fig.~\ref{fig:TLM}B). The dust shell of the progenitor model also requires large optical depths ($\tau_V\gtrsim5$) given the high mid-IR fluxes and deep $r$ and $i$-band limits. 

Figure~\ref{fig:Dmod}B shows that the post-outburst model is almost an order of magnitude more luminous ($L_*=1.1\times10^5$ $\mathrm{L}_\odot$) than the progenitor SED and is consistent with the estimated late-time optical luminosity (Fig.~\ref{fig:BVRILC}B). Post-outburst observations of the optical emission indicate that the surrounding dust is more diffuse than the optically thick progenitor shell. Model fits to the post-outburst SED require an optical depth of $\tau_V\sim0.8$. Excess emission at the $r$-band compared to the $g$ and $i$-band fluxes, which are reasonably well-fit by the model, provides evidence for strong $H\alpha$ line emission. The slope of the model SED in the near-IR is slightly steeper than the apparent slope of the $JHK_s$ fluxes, which may be due to dust temperature changes that occurred within the $\sim40$ days between the optical/near-IR and mid-IR observations. 


Both progenitor and post-outburst models indicate similar limits on the inner dust radii of $\sim3-4\times10^{14}$ cm ($\sim20-27$ AU). The estimated equilibrium temperature radii at days -2379 and 1590 are less than $\sim1/2$ of $r_\mathrm{in}$ (Fig.~\ref{fig:TLM}D); however, this is not unexpected since $r_\mathrm{in}$ is not well constrained due to degeneracies between $Y$ and $T_{D,\mathrm{in}}$. Notably the upper limit of the total shell mass in the progenitor model is consistent with the mid-IR derived dust mass assuming a gas-to-dust ratio of 100 (See Tab.~\ref{tab:Flux}). The shell mass limit from the post-outburst model is a factor of $\sim2$ lower than the mass implied from the mid-IR analysis; however, this may be attributed variations in the dust heating within the $\sim40$ days where the mid-IR and optical/near-IR photometry were acquired. The mass discrepancy may also indicate that the post-outburst circumstellar material is in a different geometry than that of the progenitor.



\section{Discussion}

\subsection{Mid-IR Dust Evolution of SN~2010da}

\subsubsection{Evaporation from the Optical Flash}

The significant decrease in reddening and optical depth towards SN~2010da during and after the outburst (Brown 2010; Bond 2010) imply the destruction of the circumstellar dust that heavily enshrouded the progenitor. The decrease in the derived mass between the progenitor and post-outburst DUSTY models (Tab.~\ref{tab:DUSTY}) is suggestive of dust destruction as well. Given the luminosity of the optical flash from the initial outburst ($L\approx3\times10^6$ $\mathrm{L}_\odot$), we can estimate the radius, $r_\mathrm{sub}$, out to where the optical/UV flux will evaporate dust by heating it beyond the sublimation temperature ($T_\mathrm{sub}\sim1700$ K; e.g. Gall et al. 2011). The sublimation radius can be estimated from Eq.~\ref{eq:temp} to be $r_\mathrm{sub}\approx 4\times10^{14}$ cm ($\sim27$ AU). Since $r_\mathrm{sub}$ is greater than the pre-outburst equilibrium temperature radius shown in Fig.~\ref{fig:TLM}D, we conclude that a significant amount of circumstellar dust was evaporated in the initial optical flash. The lower post-outburst dust mass estimates at early and intermediate times relative to the progenitor dust mass despite higher post-outburst luminosities strengthens this claim.





\subsubsection{Post-Outburst Re-Brightening: Evidence of Progenitor Survival and On-going Dust Formation}

The prominent post-outburst re-brightening across IR, optical, and X-ray wavelengths at around day $\sim1600$ (Fig.~\ref{fig:SN2010daLC}) implicates the survival of the progenitor. However, we must consider the alternative explanation where the re-brightening arises from interactions between a shock driven by the initial outburst with shells or clumps of circumstellar material. We claim that this is unlikely due to the hardness variability of the post-outburst X-ray emission reported by Binder et al. (2016): high luminosity measurements taken on days 120 and 1635 exhibit harder X-ray emission than the low luminosity emission observed on day 1450 (see Fig.~\ref{fig:SN2010daLC}C). The hardening and increasing X-ray energetics between days 1450 and 1635 are indicative of continued outburst activity from a surviving central system. Such variability is difficult to reconcile with a terminal outburst. Given the apparent survival of the progenitor, we propose that the late time (days $\sim1500-2100$) re-brightening of the IR luminosity by over a factor of 2 times the progenitor (Fig.~\ref{fig:TLM}B) is due to an increase in the luminosity of the central system and on-going dust formation.


At intermediate times (days $\sim460-1390$), the dust temperature exhibits a similar increasing trend as the IR luminosity, whereas the late-time temperature shows a sudden drop while the luminosity continues to increase until day $\sim2000$. The intermediate-time temperature-luminosity correlation is consistent with the scenario where a fixed quantity of emitting dust is at a constant distance (i.e. the dust formation radius) from a heating source increasing in brightness: $T_\mathrm{D}\propto L^{1/4}$ (Eq.~\ref{eq:temp}). Constant dust masses estimated at intermediate times (Fig.~\ref{fig:TLM}C and D) strengthen this interpretation. Ultimately, we are not sensitive to changes in the luminosity from possible warm and/or cold dust distributions and therefore require far-IR photometry to conclusively claim that the heating source is increasing in luminosity. However, it is apparent from the evolution of the mid-IR dust properties and variability of the temperature and luminosity of the optical counterpart (Fig.~\ref{fig:BVRILT}) that the energetics of the system is significantly changing many years after the initial eruption. 


Mid-IR emission observed after the outburst must be associated with dust that either survived the initial optical/UV flash, reformed in the supersaturated vapor of the previously evaporated grains, and/or reformed in the outflow from the central source. Alternatively, the emitting dust could be interstellar in nature and illuminated by the optical flash (i.e. an ``IR echo"). We first test the IR echo scenario by using Eq.~\ref{eq:temp} to estimate the predicted temperature of dust at a distance $c\,t/2$, which is where the hottest dust is located if heated by the optical/UV flash (e.g. Fox et al. 2011, 2015). At $t\approx460$ d and given an optical/UV outburst luminosity of $L\approx3\times10^6$ $\mathrm{L}_\odot$, the predicted dust temperature is $43$ K, significantly less than the temperature estimated at day 460 ($T_\mathrm{D}\approx850\pm73$ K). It is therefore unlikely that the post-outburst mid-IR emission is due to an IR echo.

The unchanging equilibrium temperature radius of the emitting dust at intermediate times ($\sim8$ AU, Fig.~\ref{fig:TLM}D) over $\sim2.5$ yr suggests that the mid-IR emission is not associated with dust reformation in the progenitor winds nor survival from the optical/UV flash. Dust formed through either channel would unlikely be located at radii that are smaller than the radii estimated for the progenitor dust, especially given an approximate shock velocity of $\sim660$ km $\mathrm{s}^{-1}$ based on the FWHM of the H$\alpha$ emission line observed from SN~2010da during the outburst (Chornock \& Berger 2010). Given a shock velocity of $660$ km $\mathrm{s}^{-1}$, the shock that would sweep up reformed or surviving dust (e.g. Kochanek 2011) will propagate $\sim300$ AU over 2.5 yr and is over an order of magnitude larger than the estimated dust radius. Although we cannot rule out the presence of surviving or reformed cold dust due to the lack of far-IR measurements, we claim that the post-outburst emission in the mid-IR is primarily associated with newly-formed dust from the surviving central system.

\subsection{On the Nature of SN~2010da}
\label{sec:discussion}

The heavily dust-enshrouded nature of the progenitor, and the low peak visual luminosity compared to core-collapse SNe, seem at first to suggest that SN\,2010da was an intermediate-luminosity transient similar to NGC~300 OT2008-1 (Bond et al.\,2009; Humphreys et al.\ 2011) and SN\,2008S (Thompson et al.\ 2009; Kochanek 2011). However, the post-outburst evolution of SN~2010da is significantly different from those of SN~2008S and NGC~300 OT2008-1, both of which are prototypes of this class of transient. One of the most prevalent interpretations reported in ATels is that SN~2010da is an outburst from a dust-enshrouded LBV (e.g. Khan et al. 2010, Elias-Rosa et al. 2010, Chornock \& Berger 2010). A confounding factor in the SN~2008S-like transient and LBV eruption interpretations is the high 0.3-10 keV X-ray luminosity of $L_x\sim4.5\times10^{38}$ erg $\mathrm{s}^{-1}$ detected within hours of the optical detection (Immler et al. 2010) and recurring, less-luminous ($L_x\sim4\times10^{37}$ erg $\mathrm{s}^{-1}$; 0.35-8 keV) outbursts (Binder et al. 2016). Based on the high X-ray emission and pre-outburst IR luminosity, Binder et al. (2016) argue that the system is a high mass X-ray binary (HMXB) where the optical companion is an LBV. We instead propose that the optical companion of the HMXB system is a supergiant (sg) B[e] star.




We discuss possible interpretations for SN~2010da given the new insight from observations of its long-term mid-IR and optical evolution. We consider the following (possibly overlapping) hypotheses for the nature of SN~2010da: a SN~2008S-like transient, an outburst from a LBV/sgB[e]-HMXB, a presumed stellar merger event like V1309~Sco (Tylenda et al. 2011), V838~Mon (Bond et al. 2003; Afsar \& Bond 2007) or the recently identified NGC 4490-OT (Smith et al. 2016), or a low-luminosity supernova explosion (e.g. electron-capture). 

\begin{figure}[t]
	\centerline{\includegraphics[scale=0.43]{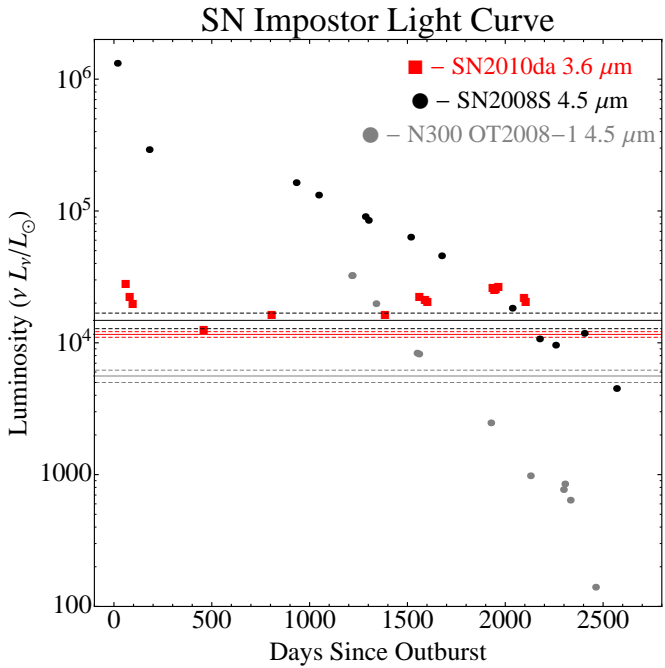}}
	\caption{Mid-IR luminosity light curve of SN~2008S and NGC~300 OT2008-1 (Adams et al. 2015) over-plotted with SN~2010da. The solid lines correspond to the progenitor flux and the dashed lines correspond to the $1-\sigma$ uncertainty in the measured flux.}
	\label{fig:ImpostorLC}
\end{figure}

\subsubsection{Comparison with SN 2008S-like Events and Low-Luminosity SNe}

The growth in IR luminosity from SN~2010da directly contrasts with the late-time mid-IR light curves of SN~2008S and OT2008-1 that decrease to over a magnitude below the mid-IR flux from the respective progenitors (Fig.~\ref{fig:ImpostorLC}). Notably, SN~2008S and OT2008-1 did not have any detections of X-ray emission, whereas the peak X-ray luminosity from SN~2010da was $L_x\sim4.5\times10^{38}$ erg $\mathrm{s}^{-1}$ (Immler et al. 2010). Subsequent X-ray and optical outbursts like those observed from SN~2010da may be hidden by substantial quantities of circumstellar dust still present around SN~2008S and OT2008-1 ($\tau_V\gtrsim100$ assuming surviving star with similar luminosity as their progenitor; Adams et al. 2015); however, this high value of $\tau_V$ is also in disagreement with the moderate optical depth estimated from the post-outburst SED of SN~2010da ($\tau_V\sim1$; Fig.~\ref{fig:Dmod}B). One may argue that differences in viewing angles or dust shell morphology may explain the non-detection of X-ray emission; however, it is difficult to invoke geometry as a means to reconcile the contrasting evolution of the late-time IR luminosity. Given the evidence we presented, we claim that SN~2010da does not fall within the class of SN~2008S-like transients.

Low-luminosity SNe have been proposed as the mechanism for 2008S-like events and may provide an alternative explanation for the SN~2010da outburst. In this scenario, SN~2010da's late-time increase in luminosity from may be due to energy input from the shock driven into the circumstellar material by the initial outburst. Interestingly, the high X-ray luminosity detected from the initial outburst is consistent with the total integrated luminosity of the system at late times (see Fig.~\ref{fig:Dmod}B and~\ref{fig:ImpostorLC}). However, the SN-interpretation is highly unlikely given the re-brightening events at mid-IR, optical, and X-ray wavelengths and the hardness variability of the X-ray emission (Binder et al. 2016) that imply the survival of the progenitor system.

\subsubsection{A Stellar Merger?}


The prospect of SN~2010da corresponding to a merger event is interesting given the increasing attention on stellar mergers (e.g. Smith et al. 2016); however, this interpretation is unlikely due to the SN~2010da's high X-ray luminosity and late-time re-brightening. Additionally, the smoothly declining optical light curve following the initial outburst of SN~2010da (Fig.~\ref{fig:BVRILC}) is not consistent with the almost periodic peaks exhibited from merger candidates V838 Mon and NGC 4490 OT. 

Interestingly, V838~Mon, the prototypical system believed to be representative of a stellar merger, exhibited a peak absolute $V$-band magnitude of about -9.8 (Sparks et al. 2008), similar to the SN~2010da peak ($M_V\sim-10.3$). Antonini et al. (2010) report variable X-ray emission from V838~Mon $\sim6$ with luminosities of $\sim10^{32-33}$ erg $\mathrm{s}^{-1}$ $\sim6$ yr after the initial outburst that is believed to be a consequence of interaction between its ejecta and early-type companion. It is difficult to reconcile the discrepancy in the X-ray energetics between V838~Mon and SN~2010da given the $\gtrsim3$ orders of magnitude difference in X-ray luminosity; therefore, we rule out the merger scenario for SN~2010da.

\subsubsection{An LBV or SgB[e] High-Mass X-ray Binary?}
\label{sec:HMXB}

The high X-ray luminosity of the initial outburst and subsequent re-brightening present strong evidence for an HMXB interpretation (Binder et al. 2016) since stellar wind-driven mechanisms have not been observed to reach such high luminosity. For example, the highest X-ray luminosity from a known LBV is $\sim10^{34}$ erg $\mathrm{s}^{-1}$ (Naz\'{e} et al. 2012), which is 4 orders of magnitude less than the peak X-ray luminosity from SN~2010da. Interestingly, optical spectra taken around the initial optical peak reveal emission lines indicative of an LBV-like outburst (e.g. H$\alpha$, He I, Fe II, and [N II]; Elias-Rosa et al. 2010; Chornock \& Berger 2010). Binder et al. (2016) therefore suggest that SN~2010da is associated with an eruption from an LBV-HMXB system. They substantiate this claim from the $\lesssim5$ Myr age estimate for the population of nearby stars inferred from the color-magnitude diagram (CMD). LBVs, however, are not found to exhibit luminosities lower than $\lesssim2.5\times10^5$ $\mathrm{L}_\odot$ (Smith, Vink, \& de Koter 2004), which is over a factor of 4 greater than the progenitor luminosity estimated from the SED model (Fig.~\ref{fig:Dmod}A). We propose that the optical companion of the HMXB is instead a sgB[e]-star.



SgB[e]-stars resemble LBVs spectroscopically and share a similar region in the HR diagram; however, sgB[e]-stars can be less luminous than LBVs and do not exhibit giant eruptions. The significant increase of the luminosity during the optical outburst ($\sim3\times10^6$ $\mathrm{L}_\odot$) is therefore the strongest evidence against the sgB[e] hypothesis. Although sgB[e] stars do not exhibit significant variability, increases in the luminosity of the optical continuum concurrent with X-ray outbursts have been observed from sgB[e]-HMXB systems. Optical observations of the April 1998 X-ray outburst ($L_\mathrm{X}\sim3\times10^{38}$ ergs $\mathrm{s}^{-1}$) of the sgB[e]-HMXB CI Cam revealed an increase in the optical luminosity by a factor of $2-5$ at the peak (Hynes et al. 2002). This is still over an order of magnitude less than the ratio of the SN~2010da progenitor and optical peak luminosity, but the CI Cam observations suggest that interactions between the sgB[e] and a compact companion can lead to an increase in the optical luminosity. 

\begin{figure}[t]
	\centerline{\includegraphics[scale=0.34]{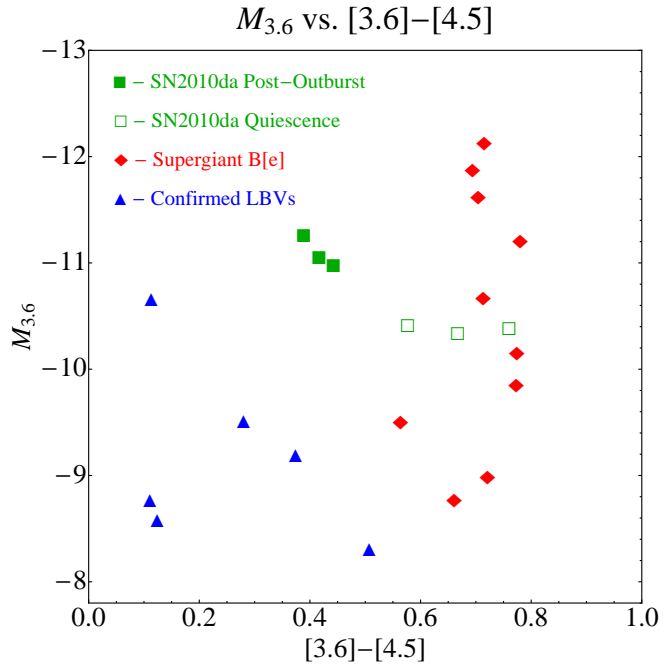}}
	\caption{Mid-IR color-magnitude diagram for sgB[e] stars and bona-fide LBVs in the Large Magellanic Cloud (Bonanos et al. 2009) over plotted with measurements of SN~2010da in ``quiescence" ($t<0$ d) and post-outburst ($t\sim2000$ d). }
	\label{fig:DLBV}
\end{figure}

A comparison of the mid-IR color and magnitudes of known sgB[e] stars and bona-fide LBVs in the Large Magellanic Cloud (Bonanos et al. 2009) to the SN~2010da progenitor (in ``quiescence") and after the eruption (``post-outburst") are shown in Fig.~\ref{fig:DLBV}. The mid-IR color and absolute magnitude of the progenitor most resembles the sgB[e] stars. Similar to the progenitor, the sgB[e] stars presented by Bonanos et al. (2009) exhibit mid-IR excesses consistent with hot $\sim600$ K circumstellar dust. Interestingly, SN~2010da migrated to bluer colors during and following the eruption, which places it in between the sgB[e] and LBV population in the color-magnitude diagram.

Although there few known, bona-fide sgB[e]-HMXB systems (CI Cam, Hynes et al. 2002; IGR J16318-4848, Filliatre \& Chaty 2004), the observed luminosities and mass in circumstellar material from literature (Filliatre \& Chaty 2004; Chaty \& Rahoui 2012; Thureau et al. 2009) are similar to what we derive for SN~2010da. From DUSTY models fit to the pre- and post-outburst SED of CI Cam, Thureau et al. (2009) derive a pre-outburst luminosity and dust mass of $2\times10^5$ $\mathrm{L}_\odot$ and $3\times10^{-6}$ $\mathrm{M}_\odot$, and a post-outburst luminosity and dust mass of $6\times10^4$ $\mathrm{L}_\odot$ and $8\times10^{-7}$ $\mathrm{M}_\odot$. Chaty \& Rahoui (2012) estimate a dust mass of $1.4\times10^{-8}$ $\mathrm{M}_\odot$ by fitting a $T_\mathrm{D}\sim 800$ disk model to the near to mid-IR spectra of IGR J16318-4848 and assuming a distance of 1.6 kpc to the system. The stellar luminosity of the optical companion at a distance of 1.6 kpc is $\sim6\times10^4$ $\mathrm{L}_\odot$. However, Chaty \& Rahoui (2012) mention the system may be further away out to $\sim4$ kpc, which would imply a dust mass of $\sim9\times10^{-8}$ $\mathrm{M}_\odot$ and a luminosity of $\sim4\times10^5$ $\mathrm{L}_\odot$. The late-time optical luminosity ($\sim10^5$ $\mathrm{L}_\odot$) and hot dust mass ($1.4\times10^{-7}$ $\mathrm{M}_\odot$) we estimate for SN~2010da are therefore comparable to that of the other sgB[e]-HMXB systems. Notably, infrared spectroscopy of IGR J16318-4848 reveal significant intrinsic optical absorption ($A_V\sim6$; Chaty \& Rahoui 2012), which is consistent with the local extinction estimated for the SN~2010da progenitor (Fig.~\ref{fig:Dmod}A). 

Based on the similar properties to other systems and the mid-IR color of the progenitor, we argue that SN~2010da is likely a sgB[e]-HMXB. Ages of sgB[e] stars are also comparable with the $\lesssim5$ Myr age inferred from the color-magnitude diagram of stars in the vicinity of SN~2010da (Binder et al. 2016). However, since there are no confirmed LBV-HMXB systems there is not enough information to conclusively rule out the LBV-HMXB hypothesis.



It is important to discuss the possible non-supergiant BeXB interpretation of SN~2010da as proposed by Binder et al. (2011) since they are more common that sgB[e]-HMXBs and also exhibit X-ray outbursts with luminosities exceeding $\sim10^{38}$ ergs $\mathrm{s}^{-1}$. The most significant difference between sgB[e] and Be stars is the lack of hot circumstellar dust around Be stars (e.g. Zickgraf et al. 1999). Given the clear presence of hot dust around SN~2010da and its progenitor, it is highly unlikely to be a non-supergiant Be star. 

\subsubsection{Orbital Constraints of the High-Mass X-ray Binary}

In this section, we assume that SN~2010da is a sgB[e]-HMXB system with a neutron star (NS) companion and derive constraints on its orbital parameters based on the observed outburst properties. High X-ray luminosity of the outbursts from sgB[e]-HMXB systems are powered by accretion onto the NS. If the NS accretes material from the sgB[e] star's winds, the large amplitude variation of the X-ray luminosity by over 2 orders of magnitude suggests that the NS is in a highly eccentric orbit. We can then place constraints on the orbital eccentricity, $e$, period, $P_\mathrm{orb}$, and semi-major axis, $a$, of the system given the measurements of the varying X-ray luminosity.  We note that the X-ray emission may also be due to mass transfer via Roche-Lobe overflow (RLOF); however, X-ray emission from HMXBs is typically dominated by accretion from stellar winds. Additionally, sgB[e] stars are not known to exhibit significant changes in radius given their unvarying spectral profile.

Under the assumption of Bondi-Hoyle accretion, the X-ray luminosity from the NS in a HMXB can be expressed as

\beq
L_\mathrm{X}=\eta \,\dot{M}_\mathrm{acc}\,c^2,
\label{eq:Lx1}
\eeq
\noindent
where $\eta$ is a constant that depends on the physics of the accretion for which we adopt a value of $\sim0.1$ (Oskinova et al. 2012), $\dot{M}_\mathrm{acc}$ is the mass accretion rate of the NS, and $c$ is the speed of light. For accretion fed by stellar winds of mass loss rate $\dot{M}_\mathrm{w}$ and velocity $v_\mathrm{w}$ regulated by the orbital motion of the NS, $\dot{M}_\mathrm{acc}$ is 

\beq
\dot{M}_\mathrm{acc}=4\pi\,\xi\frac{(G\,M_\mathrm{NS})^2}{v_\mathrm{rel}^3}\frac{\dot{M}_\mathrm{w}}{4\pi\,r^2\,v_\mathrm{w}},
\label{eq:Lx2}
\eeq
\noindent
where $\xi$ is a factor accounting for radiation pressure and the finite cooling time of the gas for which we assume is $\sim1$ (e.g. Oskinova et al. 2012), $M_\mathrm{NS}$ is the mass of the NS, $v_\mathrm{del}$ is the velocity of the NS relative to the winds, and $r$ is the orbital separation between the stellar wind source and the NS. The relative velocity is simply related to the wind velocity and the orbital velocity of the NS, $v_\mathrm{orb}$, as follows:

\beq
v_\mathrm{rel}=\sqrt{v_\mathrm{orb}^2 +v_\mathrm{w}^2}.
\label{eq:Lx3}
\eeq
\noindent
Using the \textit{vis-viva} equation, $v_\mathrm{orb}$ can be re-expressed as

\beq
v_\mathrm{orb}=\sqrt{G\,M_*\left(\frac{2}{r}-\frac{1}{a}\right)},
\label{eq:Lx4}
\eeq
\noindent
where $M_*$ is the mass of the sgB[e] star and $G$ is the gravitational constant. 

The eccentricity of the system can be constrained assuming that the observed X-ray luminosity peak, $L_\mathrm{X,p}$, occurs during periapse ($r=a(1-e)$) and that the lowest measured X-ray luminosity, $L_\mathrm{X,a}$, occurs during apoapse ($r=a(1+e)$). The ratio between the peak and minimum measured X-ray luminosity is $\sim100$ (Binder et al. 2016), which implies $L_\mathrm{X,p}/L_\mathrm{X,a}>100$. It then follows from Eq.~\ref{eq:Lx1} - \ref{eq:Lx3} that


\beq
\frac{L_\mathrm{X,p}}{L_\mathrm{X,a}}=\frac{(v_\mathrm{orb,a}^2+v_\mathrm{w}^2)^{3/2}(1+e)^2}{(v_\mathrm{orb,p}^2+v_\mathrm{w}^2)^{3/2}(1-e)^2}>100,
\label{eq:Lx5}
\eeq
\noindent
where $v_\mathrm{orb,a}$ and $v_\mathrm{orb,p}$ are the orbital velocities of the NS during apoapse and periapse, respectively. We assume that $v_\mathrm{orb}<v_w$ at apoapse and periapse, which implies

\beq
\frac{(v_\mathrm{orb,a}^2+v_\mathrm{w}^2)^{3/2}(1+e)^2}{(v_\mathrm{orb,p}^2+v_\mathrm{w}^2)^{3/2}(1-e)^2}\approx\frac{(1+e)^2}{(1-e)^2}>100.
\label{eq:Lx6}
\eeq
\noindent
The lower limit on the eccentricity is therefore

\beq
e>0.82.
\label{eq:e}
\eeq
\noindent
By combining Eqs.~\ref{eq:Lx1}-\ref{eq:Lx4}, we can constrain the semi-major axis and orbital period from the limits on the orbital eccentricity from the X-ray luminosity at periapse:

\beq
L_\mathrm{X,p}\approx0.1 \,c^2  \frac{\dot{M}_\mathrm{w}}{(a(1-e))^2\,v_\mathrm{w}}\frac{(G\,M_\mathrm{NS})^2}{\left(G\,M_*\left(\frac{2}{a(1-e)}-\frac{1}{a}\right)+v_\mathrm{w}^2\right)^{3/2}}.
\label{eq:Lx7}
\eeq
\noindent
At the late time X-ray luminosity peak at around day $\sim1600$, where $L_\mathrm{X,p}\sim4\times10^{37}$ erg $\mathrm{s}^{-1}$ (Binder et al. 2016), we infer a stellar wind mass loss rate of $\dot{M}_\mathrm{w}\sim1.4\times10^{-5}$ $\mathrm{M}_\odot$ $\mathrm{yr}^{-1}$ from our analysis in Sec.~\ref{sec:dprop} and adopting a gas-to-dust mass ratio of 100. We assume a wind velocity of $v_\mathrm{w}\sim660$ km $\mathrm{s}^{-1}$ based on the widths of the emission lines observed during the optical re-brightening at intermediate times (Chornock, Czekala, \& Berger 2011, Elias-Rosa et al. 2010, Chornock \& Berger 2010). Adopting $M_*\sim40$ $\mathrm{M}_\odot$ for the sgB[e] star and $M_\mathrm{NS}\sim1.5$ $\mathrm{M}_\odot$ for the NS, we can show from Eq.~\ref{eq:Lx7} that

\beq
a\gtrsim0.26\,\mathrm{AU}\,\,\mathrm{and}\,\,P_\mathrm{orb}\gtrsim8\,\mathrm{d}.
\eeq
\noindent
Based on the $\sim400$ d timespan between the apparent local maxima at days 1562 and 1965 in the late time mid-IR light curve (see Fig.~\ref{fig:SN2010daLC}A), we place upper limits on the orbital period and semi-major axis assuming the IR peaks correspond to the periastron passage of the NS:

\beq
P_\mathrm{orb}\lesssim400\,\mathrm{d}\,\,\mathrm{and}\,\,a\lesssim3.6\,\mathrm{AU}.
\eeq
\noindent
Known sgHMXBs typically exhibit orbital periods of 10 d or less with low eccentricities ($e\lesssim0.3$; Reig 2011) due to short circularization timescales. The long orbital periods and high eccentricities inferred for the system are therefore consistent with the Binder et al. (2016) interpretation of SN~2010da as a newly formed HMXB. Circularization timescales estimated for known sgHMXBs range from $\sim10^4-10^7$ yr (Stoyanov \& Zamanov 2009), which suggests that the presumed HMXB associated with SN~2010da is less than $10^4$ yr old. 

We can glean insight on the orbital and disk/shell configuration of the system from the upper limit derived for the semi-major axis of the binary. Estimates of the dust shell/disk radius indicate sizes greater than 7 AU (see Fig.~\ref{fig:TLM}D), which implies the disk/shell likely surrounds the entire binary system. Interactions between the sgB[e] star and compact companion may therefore be linked to the formation of the circumbinary disk/shell. Notably, the rapid increase of the observed mass of hot dust on day 1562 that occurred within $\sim100$ d of the late-time X-ray outburst (Fig.~\ref{fig:TLM}C) is consistent with this claim. It is plausible that such an outburst interacted with and swept up the older dust distribution observed at intermediate times thereby driving the material out further from the central system (day $\sim1500$ in Fig.~\ref{fig:TLM}D). Additional multi-wavelength and spectroscopic follow-up observations will be required to further substantiate this claim. 


\section{Conclusions}

In this paper, we presented observations of the optical and mid-IR outburst and recurring re-brightening events of the ``impostor" SN~2010da (Fig.~\ref{fig:SN2010da}. Both optical and mid-IR light curves reveal re-brightening events in the years following the initial outburst. We attribute the mid-IR emission to hot ($\sim700-1000$ K) dust in the vicinity of SN~2010da and use the 3.6 and 4.5 $\mu$m flux measured by \textit{Spitzer}/IRAC to estimate the evolution of the temperature, mass, luminosity, and radius of the emitting dust (Fig.~\ref{fig:TLM}). Results from this analysis revealed that the IR luminosity of the hot dust increased beyond that of the progenitor as well as a rapid increase in the observed hot dust mass at late times ($\sim1600$ days) after the outburst. We similarly use the optical light curve to study the evolution of the temperature and luminosity of the optical component (Fig.~\ref{fig:BVRILT}). These results indicated a peak optical temperature and luminosity of $\sim9500$ K and $3\times10^6$ $\mathrm{L}_\odot$, respectively, during the initial outburst. 


At pre and post-outburst epochs with sufficient multi-wavelength photometry, we fit DUSTY models to the spectral energy distributions (SEDs) in order to obtain further constraints of the dust and heating source properties. Both progenitor and post-outburst models provide parameters that are in agreement with the properties inferred from the mid-IR and optical light curve analysis. We note that the properties inferred for the models only constrain the mid-IR emitting dust that is not sensitive to colder dust that may exist at further distances from the central system.

We argue that significant quantities of pre-existing dust were destroyed via sublimation from the initial optical flash of SN~2010da by comparing the sublimation radius to the equilibrium temperature radius of the progenitor. This claim is also substantiated by the low post-outburst hot dust mass estimated at early and intermediate times relative to the progenitor dust mass. The enhanced mid-IR emission after the outburst must therefore arise from surviving, pre-existing dust, dust reformed in the winds of the progenitor, and/or newly formed dust from the outflow of the central system. We rule out surviving and reformed dust based on the the disagreement between the predicted temperatures from an IR echo of the initial optical flash and the observed hot dust temperature, and the disagreement between the expected shock radius with derived equilibrium temperature radius. Since the re-brightening emission (Fig.~\ref{fig:SN2010daLC}) and the hardness variability of the X-ray spectrum (Binder et al. 2016) implicate the survival of the progenitor, we claim that the post-outburst mid-IR emission is most likely associated with newly-formed dust from the central system. 

The nature of SN~2010da presents an interesting mystery since it exhibited an LBV-like eruption, recurring high X-ray luminosity bursts, and re-brightening mid-IR emission. Analysis of the observed properties allowed us to rule out interpretations of SN~2010da as a 2008S-like transient, a merger product, or a low-luminosity supernova. We address the leading theory of SN~2010da as a LBV-HMXB as proposed by Binder et al. (2016) and instead suggest it is a sgB[e]-HMXB. Our claim is substantiated by the low progenitor luminosity relative to known LBVs and the shared location of the progenitor in a mid-IR CMD with sgB[e] stars in the LMC as catalogued by Bonanos et al. (2009; Fig.~\ref{fig:DLBV}). Additionally, SN~2010da exhibits similar luminosities and dust masses to the two bona-fide sgB[e]-HMXB systems CI Cam and IGR J16318-4848. We, however, do not rule out a LBV-HMXB interpretation given the similarities between sgB[e] stars and LBVs, and because there are currently no confirmed LBV-HMXB systems from which to draw comparison. 

Under the sgB[e]-HMXB interpretation of SN~2010da, we were able to place constraints on its orbital parameters. Given that the X-ray emission from HMXBs typically arises from accretion of stellar winds from the optical companion onto the NS, the variations of the measured X-ray luminosity  could be used to constrain the eccentricity, orbital period, and semi-major axis of the system. Our calculations indicate a high eccentricity ($e>.82$) and long orbital period ($P_\mathrm{orb}\gtrsim8$ d) relative to known HMXBs, which supports the interpretation of SN~2010da as a newly-formed HMXB. 

Such a rare system presents unique opportunities to study how the interaction between the companions might influence the mass loss, luminosity, and temperature of the optical component as well as the orbital evolution of the system. Interestingly, upper limits we placed on the semi-major axis that were determined by the apparent periodicity of mid-IR re-brightening events indicate that the system is surrounded by a circumbinary disk/shell of hot dust. Interactions between the sgB[e] and NS at periapse may be linked to the changing properties of the surrounding dust, such as the increase in the observed hot dust mass that occurred around the same time as the recurring X-ray outburst at day $\sim1600$ (Fig.~\ref{fig:TLM}D). Follow-up observations monitoring the mid-IR, optical, and X-ray light curves will provide invaluable information on the connection between the stellar component, compact object, and circumstellar/binary dust in the system. 



\emph{Acknowledgments}. 
This work made use of observations from the \textit{Spitzer}
\textit{Space Telescope} operated by the Jet Propulsion Laboratory,
California Institute of Technology, under a contract
with NASA (PIDS 10136, 11063, \& 11053). Ground-based observations presented
were obtained from Bronberg Observatory, 
the SMARTS Consortium 1.3-m telescope at CTIO,
the 1-m Swope telescope at LCO, 
the Mount Lemmon Observing Facility, operated by the University of Minnesota,
and the Magellan Baade Telescope at LCO. This work was partially carried out at the Jet Propulsion Laboratory, California Institute of Technology, under a contract with the National Aeronautics and Space Administration. J.~J. is supported by the National Science Foundation Graduate Research Fellowship under Grant No. DGE-1144469.

R.~L. would like to thank Drew Clausen for the enlightening discussions on high-mass X-ray binaries. R.~L. also thanks Scott Adams for insightful comments.
H.~E.~B. thanks the STScI Director's Discretionary Research Fund for supporting STScI's participation in the SMARTS Consortium. We appreciate the excellent work of the CTIO/SMARTS service observers who obtained the ANDICAM images during many long clear Tololo nights: Juan Espinoza, Alberto Miranda, Mauricio Rojas, and Jacqueline Seron.

\clearpage

\begin{deluxetable}{ccc|cc|cc|cc|cc}
\tablecaption{SN~2010da Optical Discovery Magnitudes}
\tablewidth{0pt}
\tablehead{JD& $t_\mathrm{d}$ & Unfiltered Mag. & $t_\mathrm{d}$ & B & $t_\mathrm{d}$ & V & $t_\mathrm{d}$ & R & $t_\mathrm{d}$ & I}
\startdata
 2455339.67 & -3.995 & 16. &  -3.019 & 16.4 & -3.023 & 16.1 & -3.022 & 15.9  & -3.02 & 15.8 \\
 2455340.64 & -3.025 & 15.9 & -2.016 & 16.5 & -2.019 & 16.3 &  -2.018 & 16.1 & -2.017 & 15.8 \\
2455341.64 & -2.025 & 15.9 & -1.052 & 16.5 & -1.055 & 16.2 & -1.054 & 15.9 & -1.053 & 15.6 \\
2455342.61 & -1.05 & 15.9 & -0.042 & 16.2 & -0.045 & 16.4 & -0.044 & 15.8 & -0.043 & 15.4 \\
2455343.61 & -0.05 & 15.8 & 0.96 & 16.9 & 0.956 & 16.2 & 0.957 & 15.9 & 0.958 & 15.5 \\
2455343.66 & 0. & 15.7 & & & 6.97 & 16.6 & 6.971 & 16.5 & 6.972 & 16. \\
2455344.62 & 0.953 & 15.9 & & & 7.954 & 17.2 & 7.956 & 16.5 & 7.957 & 16. \\
2455350.64 & 6.974 & 16.4 & & & & & & & & \\
2455351.62 & 7.959 & 16.7  & & & & & & & & \\
2455352.62 & 8.958 & 16.7  & & & & & & & & \\
2455353.64 & 9.973 & 16.7  & & & & & & & & \\
2455355.63 & 11.966 & 16.9  & & & & & & & & \\
2455355.63 & 11.966 & 17.  & & & & & & & & \\
2455356.64 & 12.976 & 17. & & & & & & & & \\
2455357.66 & 13.996 & 17.2  & & & & & & & & \\
2455359.67 & 16.002 & 17.  & & & & & & & & \\
2455360.62 & 16.959 & 17.2  & & & & & & & & \\
2455361.62 & 17.953 & 17.4 & & & & & & & & \\
2455362.61 & 18.942 & 17.5 & & & & & & & & \\
2455363.62 & 19.956 & 17.7  & & & & & & & & \\
2455364.67 & 21.004 & 17.9  & & & & & & & & \\
2455366.64 & 22.976 & 17.8  & & & & & & & & \\
2455367.64 & 23.977 & 18.  & & & & & & & & \\
2455372.63 & 28.964 & 18.6  & & & & & & & & \\
2455381.61 & 37.941 & 19.2  & & & & & & & & \\
\enddata

\tablecomments{$t_\mathrm{d}=0$ corresponds to the date of peak optical brightness (JD = 2455343.66). The observations were performed by L. A. G. Monard at Bronberg Observatory}
	\label{tab:BMonard}
\end{deluxetable}

\clearpage
\begin{deluxetable}{cccccccccc}
\tablecaption{Optical Magnitudes, Color, and Luminosity}
\tablewidth{0pt}
\tablehead{JD & $t_\mathrm{d}$ & $B$ &  $V/g$ &$R/r$ &$I/i$ &$B-V$ &$V-I$/$g-i$ & Log($\mathrm{L}/\mathrm{L}_\odot)$ & $\mathrm{T}$ (K)}
\startdata
 2455341.86 & -1.803 & 16.69 (0.01) & 16.34 (0.01) & 15.97 (0.01) & 15.68 (0.01) & 0.36 (0.01) & 0.66 (0.01) & 6.4 (0.) & 9390 (100) \\
 2455341.87 & -1.799 & 16.7 (0.02) & 16.31 (0.02) & 15.94 (0.01) & 15.68 (0.02) & 0.38 (0.02) & 0.64 (0.02) & 6.4 (0.01) & 9360 (190) \\
 2455342.88 & -0.785 & 16.6 (0.02) & 16.24 (0.01) & 15.88 (0.03) & 15.6 (0.04) & 0.36 (0.02) & 0.64 (0.04) & 6.44 (0.) & 9470 (290) \\
 2455343.86 & 0.194 & 16.55 (0.03) & 16.17 (0.02) & 15.81 (0.01) & 15.5 (0.02) & 0.38 (0.03) & 0.66 (0.03) & 6.46 (0.02) & 9230 (240) \\
 2455346.89 & 3.224 & 16.83 (0.01) & 16.46 (0.01) & 16.02 (0.01) & 15.78 (0.01) & 0.37 (0.01) & 0.68 (0.01) & 6.34 (0.01) & 9200 (110) \\
 2455347.9 & 4.231 & 17.08 (0.02) & 16.67 (0.02) & 16.17 (0.02) & 15.98 (0.02) & 0.41 (0.03) & 0.69 (0.03) & 6.24 (0.02) & 8930 (240) \\
 2455348.87 & 5.206 & 17.27 (0.01) & 16.85 (0.01) & 16.31 (0.01) & 16.08 (0.01) & 0.42 (0.02) & 0.77 (0.02) & 6.15 (0.01) & 8460 (110) \\
 2455350.88 & 7.22 & 17.41 (0.) & - & 16.42 (0.) & -  & - & - & - & - \\
 2455350.91 & 7.247 & 17.4 (0.02) & 16.97 (0.01) & 16.41 (0.01) & 16.18 (0.02) & 0.43 (0.02) & 0.79 (0.02) & 6.1 (0.01) & 8330 (120) \\
 2455351.86 & 8.199 & 17.56 (0.01) & 17.12 (0.01) & 16.52 (0.01) & 16.28 (0.01) & 0.44 (0.01) & 0.84 (0.01) & 6.04 (0.01) & 8110 (80) \\
 2455352.87 & 9.201 & 17.64 (0.01) & 17.21 (0.01) & 16.57 (0.01) & 16.34 (0.01) & 0.43 (0.01) & 0.87 (0.01) & 6. (0.01) & 7980 (80) \\
 2455352.88 & 9.216 & 17.64 (0.01) & 17.16 (0.01) & 16.58 (0.01) & 16.36 (0.01) & 0.48 (0.01) & 0.81 (0.02) & 6.01 (0.01) & 8060 (90) \\
 2455353.9 & 10.233 & 17.8 (0.01) & 17.31 (0.01) & 16.69 (0.01) & 16.44 (0.01) & 0.48 (0.01) & 0.87 (0.01) & 5.94 (0.01) & 7780 (70) \\
 2455355.85 & 12.189 & 18. (0.01) & 17.55 (0.01) & 16.84 (0.01) & 16.63 (0.01) & 0.46 (0.02) & 0.92 (0.02) & 5.86 (0.01) & 7690 (80) \\
 2455355.87 & 12.204 & 18. (0.01) & 17.54 (0.01) & 16.86 (0.01) & 16.6 (0.02) & 0.46 (0.02) & 0.94 (0.02) & 5.86 (0.01) & 7560 (90) \\
 2455358.93 & 15.266 & 18.39 (0.01) & 17.9 (0.02) & 17.14 (0.01) & 16.94 (0.02) & 0.49 (0.02) & 0.96 (0.03) & 5.71 (0.01) & 7400 (100) \\
 2455358.94 & 15.273 & 18.4 (0.02) & 17.95 (0.01) & 17.16 (0.01) & 16.92 (0.02) & 0.45 (0.02) & 1.03 (0.03) & 5.7 (0.01) & 7250 (120) \\
 2455360.91 & 17.25 & 18.52 (0.02) & 18.04 (0.01) & 17.23 (0.01) & 17.02 (0.02) & 0.48 (0.02) & 1.02 (0.02) & 5.65 (0.01) & 7220 (80) \\
 2455364.93 & 21.264 & 18.89 (0.02) & 18.46 (0.02) & 17.53 (0.02) & 17.38 (0.02) & 0.43 (0.03) & 1.07 (0.03) & 5.5 (0.02) & 7150 (110) \\
 2455369.85 & 26.181 & 19.18 (0.03) & 18.72 (0.03) & 17.8 (0.02) & 17.61 (0.02) & 0.46 (0.04) & 1.12 (0.04) & 5.39 (0.02) & 6930 (130) \\
 2455369.86 & 26.196 & 19.18 (0.03) & 18.68 (0.03) & 17.78 (0.02) & 17.56 (0.03) & 0.49 (0.04) & 1.13 (0.05) & 5.4 (0.02) & 6810 (140) \\
 2455372.86 & 29.193 & 19.53 (0.06) & 19.02 (0.06) & 18. (0.03) & 17.89 (0.04) & 0.52 (0.09) & 1.12 (0.07) & 5.26 (0.05) & 6770 (200) \\
 2455372.87 & 29.201 & 19.45 (0.07) & 18.96 (0.04) & 18.01 (0.02) & 17.85 (0.03) & 0.49 (0.08) & 1.12 (0.05) & 5.29 (0.04) & 6850 (210) \\
 2455374.89 & 31.223 & - & - & 17.84 (0.03) & 17.72 (0.04) & - & - & - & - \\
 2455374.9 & 31.238 & - & - & 18.18 (0.05) & 18. (0.07) & - & - & - & - \\
 2455376.87 & 33.201 & 19.52 (0.06) & 19.11 (0.04) & 18.1 (0.02) & 17.93 (0.03) & 0.41 (0.07) & 1.17 (0.06) & 5.25 (0.04) & 6830 (190) \\
 2455377.95 & 34.29 & 19.62 (0.06) & 19.24 (0.11) & 18.29 (0.06) & 17.96 (0.06) & 0.38 (0.12) & 1.27 (0.12) & 5.21 (0.06) & 6570 (180) \\
 2455378.82 & 35.157 & 19.67 (0.06) & 19.18 (0.04) & 18.24 (0.02) & 18.01 (0.03) & 0.5 (0.07) & 1.16 (0.05) & 5.2 (0.04) & 6690 (170) \\
 2455380.84 & 37.178 & 19.78 (0.04) & 19.26 (0.03) & 18.29 (0.02) & 18.1 (0.03) & 0.52 (0.05) & 1.15 (0.04) & 5.17 (0.03) & 6670 (140) \\
 2455380.86 & 37.193 & 19.7 (0.04) & 19.3 (0.04) & 18.28 (0.03) & 18.15 (0.04) & 0.4 (0.06) & 1.15 (0.06) & 5.17 (0.04) & 6960 (210) \\
 2455382.79 & 39.121 & 19.8 (0.04) & 19.35 (0.03) & 18.36 (0.02) & - & 0.45 (0.06) & - & - & - \\
 2455385.9 & 42.239 & 20. (0.03) & 19.49 (0.02) & 18.52 (0.02) & - & 0.51 (0.04) & - & - & - \\
 2455385.92 & 42.252 & 20.05 (0.04) & 19.56 (0.05) & 18.53 (0.03) & 18.34 (0.04) & 0.49 (0.06) & 1.22 (0.06) & 5.06 (0.03) & 6530 (140) \\
 2455388.92 & 45.26 & 20.15 (0.03) & 19.63 (0.03) & 18.64 (0.02) & - & 0.51 (0.04) & - & - & - \\
 2455390.86 & 47.197 & 20.22 (0.02) & 19.74 (0.02) & 18.72 (0.02) & 18.51 (0.02) & 0.48 (0.03) & 1.24 (0.03) & 4.99 (0.02) & 6490 (90) \\
 2455410.84 & 67.176 & - & - & 19.17 (0.04) & 18.88 (0.04) & - & - & - & - \\
 2455417.87 & 74.205 & - & - & 19.39 (0.04) & 19.06 (0.04) & - & - & - & - \\
 2455421.79 & 78.126 & 20.83 (0.04) & 20.38 (0.04) & 19.33 (0.04) & 19.06 (0.04) & 0.45 (0.06) & 1.32 (0.06) & 4.74 (0.03) & 6270 (130) \\
 2455425.78 & 82.118 & 21.07 (0.05) & 20.64 (0.04) & 19.52 (0.04) & 19.18 (0.04) & 0.43 (0.06) & 1.45 (0.06) & 4.66 (0.03) & 5940 (130) \\
 2455430.81 & 87.15 & - & - & 19.52 (0.05) & 19.37 (0.06) & - & - & - & - \\
 2455444.82 & 101.16 & 21.3 (0.06) & 20.83 (0.05) & 19.79 (0.05) & 19.49 (0.06) & 0.47 (0.07) & 1.34 (0.07) & 4.56 (0.04) & 6180 (190) \\
 2455877.62 & 533.956 & 19.84 (0.06) & 19.27 (0.04) & 18.5 (0.04) & 18.15 (0.03) & 0.58 (0.07) & 1.12 (0.05) & 5.15 (0.04) & 6660 (150) \\
 2455885.63 & 541.965 & 20.35 (0.03) & 19.81 (0.03) & 18.89 (0.03) & 18.49 (0.03) & 0.54 (0.05) & 1.32 (0.04) & 4.96 (0.02) & 6120 (100) \\
 2455893.59 & 549.927 & 20.78 (0.05) & 20.22 (0.04) & 19.33 (0.04) & 18.9 (0.04) & 0.56 (0.06) & 1.32 (0.06) & 4.79 (0.03) & 6080 (130) \\
 2455895.55 & 551.89 & 21.28 (0.07) & 20.39 (0.05) & - & - & 0.89 (0.09) & - & - & - \\
 2455896.61 & 552.944 & 21.34 (0.08) & 20.38 (0.06) & - & - & 0.96 (0.1) & - & - & - \\
 2455900.62 & 556.956 & - & - & 19.83 (0.09) & 19.13 (0.07) & - & - & - & - \\
 2455909.6 & 565.939 & - & - & 20.09 (0.1) & 19.4 (0.08) & - & - & - & - \\
 2455916.6 & 572.939 & - & - & 20.18 (0.09) & 19.43 (0.08) & - & - & - & - \\
 
 2456794.94 & 1451.273 & - & - & 19.99 (0.17) & 20.11 (0.21) & - & - & - & - \\
 2456828.94 & 1485.279 & - & 21.08 (0.17) & 19.73 (0.08) & - & - & - & - & - \\
 2456828.94 & 1485.28 & - & 20.97 (0.17) & 19.72 (0.08) & 19.76 (0.13) & - & 1.22 (0.21) & 4.74 (0.18) & 6640 (1100) \\
 2456829.89 & 1486.23 & - & 20.89 (0.14) & 19.68 (0.07) & 19.74 (0.14) & - & 1.15 (0.2) & 4.78 (0.19) & 6920 (1130) \\
 2456870.79 & 1527.129 & - & 20.42 (0.08) & 19.47 (0.06) & 19.51 (0.09) & - & 0.91 (0.12) & 4.96 (0.22) & 8150 (980) \\
 2456870.79 & 1527.13 & - & 20.53 (0.09) & 19.46 (0.06) & 19.44 (0.08) & - & 1.09 (0.12) & 4.92 (0.2) & 7170 (760) \\
 2456871.85 & 1528.186 & - & 20.67 (0.1) & 19.54 (0.06) & 19.72 (0.1) & - & 0.96 (0.14) & 4.86 (0.22) & 7860 (1080) \\
 2456871.85 & 1528.187 & - & 20.83 (0.11) & 19.59 (0.06) & 19.85 (0.11) & - & 0.98 (0.15) & 4.8 (0.22) & 7730 (1110) \\
 2456980.56 & 1636.895 & - & 20.74 (0.09) & - & - & - & - & - & - \\
 2456980.56 & 1636.896 & - & 20.46 (0.1) & 19.51 (0.06) & 19.46 (0.08) & - & 1. (0.13) & 4.94 (0.22) & 7620 (960) \\
 2457011.58 & 1667.913 & - & 20.74 (0.1) & 19.58 (0.06) & 19.51 (0.09) & - & 1.23 (0.13) & 4.83 (0.17) & 6570 (710) \\
 2457011.58 & 1667.915 & - & 20.72 (0.1) & 19.49 (0.06) & 19.58 (0.09) & - & 1.14 (0.13) & 4.84 (0.19) & 6950 (790) \\
 2457176.91 & 1833.243 & - & 20.71 (0.23) & 19.05 (0.08) & 19.22 (0.1) & - & 1.49 (0.25) & 4.85 (0.09) & 5730 (930) \\
 2457176.91 & 1833.244 & - & 20.39 (0.15) & 19.24 (0.08) & 19.36 (0.11) & - & 1.03 (0.18) & 4.98 (0.22) & 7490 (1220) \\
 2457222.82 & 1879.153 & - & 20.43 (0.08) & 19.34 (0.05) & 19.32 (0.07) & - & 1.1 (0.11) & 4.96 (0.2) & 7120 (710) \\
 2457222.82 & 1879.155 & - & 20.37 (0.08) & 19.27 (0.05) & 19.27 (0.08) & - & 1.1 (0.11) & 4.98 (0.2) & 7130 (680) \\
 2457247.88 & 1904.217 & - & 20.44 (0.08) & 19.22 (0.05) & 19.35 (0.08) & - & 1.09 (0.12) & 4.95 (0.2) & 7170 (740) \\
 2457247.88 & 1904.219 & - & 20.36 (0.07) & 19.32 (0.05) & 19.23 (0.07) & - & 1.13 (0.09) & 4.99 (0.19) & 7020 (590) \\
 2457339.67 & 1996.006 & - & 20.58 (0.1) & 19.37 (0.06) & 19.4 (0.08) & - & 1.18 (0.13) & 4.9 (0.19) & 6800 (730) \\
 2457339.67 & 1996.008 & - & 20.49 (0.09) & 19.38 (0.05) & 19.5 (0.09) & - & 0.99 (0.13) & 4.94 (0.22) & 7690 (930) \\
\enddata

\tablecomments{Numbers in the parentheses correspond to the 1-$\sigma$ uncertainty of the adjacent value. Entries with ``0" indicate errors less than $<0.01$. Empty entries (``-") designate observations where SN~2010da was not detected, or the data are bad/non-existent.  $t_\mathrm{d}$=0 corresponds to the date of peak optical brightness from the optical discovery light curve (JD = 2455343.66).}
	\label{tab:FluxO}
\end{deluxetable}

\clearpage

\begin{deluxetable}{cccccccc}
\tablecaption{Mid-IR Flux and Dust Properties}
\tablewidth{0pt}
\tablehead{JD &  $t_\mathrm{d}$ & $F_\mathrm{3.6}$ ($\mu$Jy)& $F_\mathrm{4.5}$ ($\mu$Jy) &$T_\mathrm{d}$ (K) &$L_\mathrm{IR}$ $(\times\,10^4\,\mathrm{L}_\odot)$ &$M_{d}$ $(\times\,10^{-7}\,\mathrm{M}_\odot)$ & $\mathrm{R}_\mathrm{eq}$ (AU)}
\startdata
 2452964.56 & -2379.1 & 112.31 (5.63) & 144.65 (5.68) & 646 (48) & 1.4 (0.1) & 3.84 (1.4) & 12.2 (2.) \\
 2454463.38 & -880.286 & 115.05 (5.54) & 125.17 (5.15) & 751 (66) & 1.4 (0.2) & 1.65 (0.6) & 8.9 (1.7) \\
 2454463.44 & -880.223 & 107.91 (5.45) & 127.65 (5.25) & 695 (57) & 1.3 (0.1) & 2.38 (0.9) & 10.1 (1.8) \\
2455187.38 & -156.287 & 168.96 (3.62) & - & - & - & - & - \\
 2455210.19 & -133.476 & 271.02 (4.6) & - & - & - & - & - \\
 2455404.68 & 61.0151 & 265.08 (4.72) & - & - & - & - & - \\
 2455424.6 & 80.9322 & 211.54 (4.74) & - & - & - & - & - \\
 2455440.26 & 96.5943 & 186.65 (4.53) & - & - & - & - & - \\
 2455803.21 & 459.542 & 120.14 (4.77) & 115.83 (4.15) & 852 (73) & 1.5 (0.3) & 0.91 (0.3) & 7.3 (1.4) \\
 2456150.05 & 806.389 & 153.59 (5.04) & 134.81 (4.33) & 953 (80) & 2.2 (0.5) & 0.71 (0.2) & 6.9 (1.3) \\
 2456730.01 & 1386.34 & 155.56 (4.36) & 129.88 (4.35) & 1019 (88) & 2.5 (0.6) & 0.54 (0.1) & 6.4 (1.2) \\
 2456905.74 & 1562.08 & 211.75 (4.86) & 193.13 (4.16) & 909 (49) & 2.9 (0.3) & 1.2 (0.2) & 8.7 (1.) \\
 2456933.92 & 1590.25 & 200.98 (5.09) & 187.02 (3.98) & 887 (49) & 2.7 (0.3) & 1.27 (0.2) & 8.8 (1.) \\
 2456945.26 & 1601.6 & 193.37 (4.01) & 174.45 (3.47) & 921 (46) & 2.7 (0.3) & 1.03 (0.2) & 8.2 (0.9) \\
 2457279.93 & 1936.26 & 249.64 (4.24) & 221.77 (3.58) & 939 (39) & 3.5 (0.3) & 1.22 (0.2) & 9. (0.8) \\
 2457287.84 & 1944.17 & 237.62 (5.13) & 220.15 (4.15) & 892 (43) & 3.2 (0.3) & 1.46 (0.2) & 9.5 (1.) \\
 2457295.22 & 1951.55 & 243.65 (5.29) & 221.92 (3.73) & 910 (43) & 3.3 (0.3) & 1.37 (0.2) & 9.3 (0.9) \\
 2457309.06 & 1965.4 & 251.61 (5.18) & 230.2 (3.7) & 905 (40) & 3.4 (0.3) & 1.45 (0.2) & 9.6 (0.9) \\
 2457440.6 & 2096.94 & 209.07 (5.11) & 196.29 (4.63) & 878 (49) & 2.8 (0.3) & 1.38 (0.3) & 9.1 (1.1) \\
 2457448.08 & 2104.42 & 194.91 (5.18) & 187.36 (4.6) & 855 (49) & 2.5 (0.3) & 1.46 (0.3) & 9.2 (1.1) \\
\enddata

\tablecomments{$F_\mathrm{3.6}$ and $F_\mathrm{3.6}$ are the 3.6 and 4.5 $\mu$m fluxes of SN~2010da, respectively. $T_\mathrm{d}$, $M_\mathrm{d}$, and $\mathrm{R}_\mathrm{eq}$ are the dust color temperature, dust mass, and equilibrium temperature radius derived from $F_\mathrm{3.6}$ and $F_\mathrm{4.5}$. Values in the parentheses indicate the 1-$\sigma$ uncertainty of the adjacent value. SN~2010da was not observed between days -156.3 and 96.6 at 4.5 $\mu$m.}
	\label{tab:Flux}
\end{deluxetable}
\clearpage

\begin{deluxetable}{ccccccc}
\tablecaption{DUSTY Model Constraints on Circumstellar Dust}
\tablewidth{0pt}
\tablehead{ Epoch & $T_\mathrm{in}$ (K) &  $L_*$ $(\times10^4\,\mathrm{L}_\odot)$ &$\tau_V$ &$Y$ &$r_\mathrm{in}$ $(\times10^{14}$ cm) &$M_e$ $(\times10^{-5}\,\mathrm{M}_\odot)$  }

\startdata
	Progenitor & $\gtrsim1200$ & 1.6 & $\gtrsim5$ & $\gtrsim1$& $\lesssim3.9$ & $\lesssim4.8$\\
	Post-Outburst & $\gtrsim1800$ & 11 & $\sim0.8$ & $\gtrsim5$& $\lesssim2.7$ & $\lesssim0.4$\\
\enddata

\tablecomments{$T_\mathrm{in}$ is the dust temperature at the inner radius of the shell, $r_\mathrm{in}$, $L_*$ is the luminosity of the heating source, $\tau_\mathrm{V}$ is the optical depth through the dust shell, $Y$ is the geometrical thickness as a multiplicative factor of $r_\mathrm{in}$, and $M_\mathrm{e}$ is the total mass of the shell derived from Eq.~\ref{eq:mass}.}
	\label{tab:DUSTY}
\end{deluxetable}

\clearpage

\end{document}